\documentclass[useAMS,usenatbib]{mn2e}
\usepackage{graphicx,amssymb,color}
\usepackage[normalem]{ulem}

\title[Tides between planets and white dwarfs]
{Orbital relaxation and excitation of planets tidally interacting with white dwarfs}

\author[]{Dimitri Veras$^{1,2}$\thanks{E-mail: d.veras@warwick.ac.uk}\thanks{STFC Ernest Rutherford Fellow},
Michael Efroimsky$^3$,
Valeri V. Makarov$^3$,
Gwena\"{e}l Bou\'{e}$^4$,
\newauthor
Vera Wolthoff$^5$,
Sabine Reffert$^5$,
Andreas Quirrenbach$^5$,
Pier-Emmanuel Tremblay$^2$,
\newauthor
Boris T. G\"{a}nsicke$^{1,2}$
\\
$^{1}$Centre for Exoplanets and Habitability, University of Warwick, Coventry CV4 7AL, UK
\\
$^{2}$Department of Physics, University of Warwick, Coventry CV4 7AL, UK
\\
$^{3}$US Naval Observatory, Washington DC 20392, USA
\\
$^{4}$IMCCE, Observatoire de Paris, UPMC Univ. Paris 6, PSL Research University, Paris, France
\\
$^{5}$Landessternwarte, Zentrum f\"{u}r Astronomie der Universit\"{a}t Heidelberg, K\"{o}nigstuhl 12, 69117 Heidelberg, Germany
}

\pubyear{2019}

\begin{document}
\label{firstpage}
\pagerange{\pageref{firstpage}--\pageref{lastpage}}
\maketitle

\begin{abstract}
Observational evidence of white dwarf planetary systems is dominated by the remains of exo-asteroids through accreted metals, debris discs, and orbiting planetesimals. However, exo-planets in these systems play crucial roles as perturbing agents, and can themselves be perturbed close to the white dwarf Roche radius. Here, we illustrate a procedure for computing the tidal interaction between a white dwarf and a near-spherical solid planet. This method determines the planet's inward and/or outward drift, and whether the planet will reach the Roche radius and be destroyed.  We avoid constant tidal lag formulations and instead employ the self-consistent secular Darwin-Kaula expansions from \cite{bouefr2019}, which feature an arbitrary frequency dependence on the quality functions. We adopt wide ranges of dynamic viscosities and spin rates for the planet in order to straddle many possible outcomes, and provide a foundation for the future study of individual systems with known or assumed rheologies. We find that: (i) massive Super-Earths are destroyed more readily than minor planets (such as the ones orbiting WD 1145+017 and SDSS J1228+1040), (ii) low-viscosity planets are destroyed more easily than high-viscosity planets, and (iii) the boundary between survival and destruction is likely to be fractal and chaotic.  
\end{abstract}


\section{Introduction}

The discovery of bona fide asteroids surrounding two white dwarfs on low-to-moderate
eccentric orbits \citep{vanetal2015,manetal2019} has refocused efforts to understand
the fate of planetary systems.  Debris from destroyed planetesimals both surround and 
accrete onto the white dwarf. This destruction is preceded by the close gravitational 
interaction between a white dwarf
and a planetary body, a process which has been investigated only sparsely.  
Hence, the near-complete dearth of dedicated investigations into tidal effects between 
white dwarfs and planets necessitate a remedy; this paper represents an initial step 
towards addressing this deficiency.

\subsection{Planetary system evolution}

When a main-sequence exo-planet host becomes a giant branch star, violent changes ensue,
transforming the planetary system \citep{veras2016}. The star loses most of its mass,
expanding the orbits of surviving planets, moons, asteroids and comets
\citep{omarov1962,hadjidemetriou1963,veretal2011,adaetal2013,veretal2013a}.
This expansion may trigger gravitational instabilities \citep{debsig2002} -- even without
the presence of any stellar companions -- with
important consequences for destruction, escape and, in general, orbital rearrangement
\citep{bonetal2011,debetal2012,portegieszwart2013,musetal2013,veretal2013b,voyetal2013,
frehan2014,musetal2014,veretal2016,musetal2018,smaetal2018}.

Further, giant branch stars exhibit luminosities which exceed the Sun's by several orders
of magnitude. Hence, pebbles and dust are propelled and dragged at different rates and can also
cross stability boundaries with planets \citep{donetal2010,veretal2015b}.
The consequences for asteroids are perhaps more dramatic: they can be spun up
to the point of rotational fission \citep{veretal2014c} through the
YORP (Yarkvosky-O'Keefe-Radzievskii-Paddack) effect. The survivors
can be thrust inward or outward, speeding past planets, due to a supercharged
Yarkovsky effect \citep{veretal2015b,veretal2019}. The end result is a widely dispersed
debris field ranging from pebbles to planets at distances from a few au to hundreds
or thousands of au.

The inner boundary of this debris is set by the physical expansion of the giant branch star itself,
which can reach several au in size. Tidal effects could even draw in and engulf planets which are further away
than the stellar radius. This critical tidal engulfment distance has been the subject of much interest and investigation
\citep{villiv2009,kunetal2011,musvil2012,adablo2013,norspi2013,valras2014,viletal2014,
madetal2016,staetal2016,galetal2017,raoetal2018}. The actual destructive spiral-in process and timescale for
planets within the stellar envelope has also been estimated \citep{jiaspr2018,macetal2018}, as well
as the consequences for the giant star \citep{neltau1998,massarotti2008,caretal2012,caretal2013}.

The star eventually transitions into a white dwarf, which is comparable in size to the Earth but has 
a Roche radius which extends outward to about one Solar radius for a wide range of orbiting
secondary structures \citep{veretal2017}. Within this Roche radius observations are abundant.
Metallic debris is seen in the white dwarf photosphere 
\citep[e.g.][]{zucetal2007,gaeetal2012,juryou2014,koeetal2014,haretal2018,holetal2018}
and just outside is the presence of Solar radii-scale debris discs \citep{farihi2016}.

Perhaps most eye-catching are the discoveries of minor planets orbiting white dwarfs.
At least one planetesimal is currently disintegrating around WD 1145+017 
with an orbital period of about 4.5 hours \citep{vanetal2015} and one other 
planetesimal is intact and embedded inside the disc surrounding 
SDSS J1228+1040 with an orbital period of about 2.06 hours \citep{manetal2019}. 
The orbits of the disintegrating planetesimals are thought to be nearly
circular \citep{guretal2017,veretal2017} whereas the embedded planetesimal's
eccentricity is likely to be many tenths (potentially 0.53), and accompanies a disc-based intensity
pattern with a likely eccentricity of 0.4 \citep{manetal2019}.

How objects can be emplaced at one Solar radius from several au -- and then circularised --
is still subject to debate. Although dust
particles can be drawn into the white dwarf
through Poynting-Robertson drag, larger objects need to be gravitationally perturbed onto
highly eccentric ($>0.99$) orbits. In order to reach the white dwarf's Roche radius,
\cite{graetal1990} and \cite{jura2003} proposed that these objects are perturbed there
(from distances of a least a few au) and then break up into discs, an idea which has been realised through numerical
simulations \citep{debetal2012,veretal2014a,veretal2015a,maletal2019} and purely analytic 
formulations \citep{wyaetal2014,broetal2017,kenbro2017a,kenbro2017b}. \cite{veretal2018b}
argued extensively that the presence of a planet is a near-necessity to perturb asteroids close to the white 
dwarf in a single-star system.

Nevertheless, the major planets themselves can be perturbed close to and onto the white dwarf,
particularly when the planets are comparable in mass to one another \citep{vergae2015,veretal2016}; a deficiency
with those two studies is the absence of tidal modelling, which can introduce uncertainties in the orbital behaviour
of a planet passing close
to the white dwarf. These planets would play major roles in perturbing smaller objects
and dust onto the star, with implications
for fragmentation, sublimation and the resulting debris size distributions \citep{wyaetal2014,broetal2017}.

Additional motivations for studying tides between a white dwarf and planet include the prospect
of the formation of second- or third-generation planets, which may even arise from second- or third-generation
debris discs \citep{perets2011,schdre2014,voletal2014,hogetal2018,vanetal2018}. Such planets would
reside on near-circular orbits, and may be out of reach of detection. Further, the destruction of a planet,
either through collisions with other planets or by intersecting the white dwarf Roche radius, 
could result in easily detectable events \citep{beasok2013,vanetal2015} and smaller planets
orbiting near the white dwarf. Finally, moons of planets can be stripped, perturbed towards the white
dwarf, and as a result interact with the white dwarf in close proximity \citep{payetal2016,payetal2017}.

In the year 2018, the number of known white dwarfs increased by an order of magnitude \citep{genetal2019}.
An accompanying increase in the number of previously known white dwarf planetary systems
\citep{kleetal2013,kepetal2015,kepetal2016,holetal2017} is likely to follow with the next generation of spectroscopic
initiatives and other missions. This current study may be particularly impactful if planets are discovered orbiting white dwarfs
either spectroscopically from ground-based facilities, or through TESS, 
LSST \citep{corkip2018,lunetal2018}, Gaia \citep{peretal2014} and LISA \citep{steetal2018,tamdan2018}.
Whether tidal effects had or will have a role in these systems can then be estimated through orbital proxies
and stability analyses \citep{veretal2018a}.

 \subsection{Tidal formulations\label{formulations}}

 Despite these numerous motivations, tidal investigations of white dwarfs have been limited to star-star interactions \citep{fullai2011,fullai2012,fullai2013,fullai2014,valetal2012,sraetal2014,vicetal2017,mcnetal2019}.
 In contrast, abundant studies of the main-sequence and giant-branch phases of evolution have analysed the star-planet interaction, and have adopted a variety of approaches with both equilibrium and dynamical tides.

 Two commonly-used star-planet tidal models are the CPL (constant phase lag) model pioneered by \citet{macdonald1964} and \citet{goldreich1966}, and the CTL (constant time lag) model suggested in \citet{hut1981} and \citet{eggetal1998}. Introduced for the ease of analytical treatment and for illustrative purposes, these models have very limited application in quantitative analysis. For example, the CPL model is both physically and mathematically inconsistent \citep{efrmak2013}. The CTL model may (arguably) render a tolerable approximation for stars, but its application to terrestrial bodies is possible only in the situations when these bodies are hot and plastic \citep{makarov2015}. Under normal conditions, these models are applicable neither to solids, nor to solids with partial melt.

 Generally, employment of an incorrect tidal model results in distortions of the true tidal evolution timescales and fails to 
describe correctly the process of tidal capture in spin-orbit resonances
 \citep[for details, see][]{noyetal2014}. Also, the use of simplistic models sometimes leads to qualitatively wrong conclusions, like with the emergence of the so-called pseudo-synchronous rotation state \citep{makefr2013}. Perhaps most importantly, the use of an incorrect tidal model prevents the correct assessment of the tidal heat produced in a body.

Although some of the hitherto published far-reaching conclusions on tidal dynamics of celestial bodies were based on simplistic {\it{ad hoc}} models, here we wish to avoid such flaws. Hence, we rely on a versatile formalism that can accurately reflect the increasing detail with which the community has been characterising planets and stars.

In order to accommodate such detail, consistent analytical modeling of linear bodily tides should comprise two steps: (1) First, Fourier-expansion of both the perturbing potential and the induced incremental potential of the tidally perturbed body. Each of these two expansions is an infinite sum over the tidal Fourier modes $\omega_{lmpq}$ which are numbered with aid of some integers $\,l,\,m,\,p,\,q\,$ (so that, technically, each such expansion becomes a sum over $l,\,m,\,p,\,q$). Appendix C describes the nature of these indices in more detail. (2) Second, linking each Fourier component of the incremental tidal potential to a corresponding Fourier component of the perturbing potential. This link implies establishing~---~for each Fourier mode $\omega_{lmpq}$~---~both the phase lag $\epsilon_l(\omega_{lmpq})$ and the ratio of the magnitudes, termed the dynamical Love number and denoted with $k_l(\omega_{lmpq})$. Owing to the interplay of rheology and self-gravitation, the phase lags and Love numbers possess nontrivial frequency dependencies; see Section \ref{link} below.

 The construction of a consistent mathematical theory of bodily tides was begun by \citet{darwin1879} who wrote down partial sums of the Fourier expansions of both the perturbing potential and the incremental potential of a tidally perturbed sphere. A great development of this theory was achieved by \citet{kaula1964} who managed to derive complete series expansions for these potentials. The paper by \citet{efrmak2013} offers a relatively simple introduction to the Darwin-Kaula theory and also explains how tidal friction and lagging should be built into that theory.

 With these tools at hand, it is possible to develop similar Fourier-type expansions for the orbital elements' rates \citep{kaula1964}
 and the tidal torque \citep{efroimsky2012a}. Fortuitously, \cite{bouefr2019} recently executed a complete rederivation of the expressions for the Keplerian elements' tidal rates. They pointed out a minor omission in Kaula's old treatment, and also explained that Kaula's expression for the inclination rate contained a more important flaw. We use the \cite{bouefr2019} formulae in this paper. For readability purposes for a wide audience, these formulae and a few others have been relegated to Appendices B-C.

 Computation of these equations, however, pose a challenge. As the eccentricity of the orbit approaches unity, the number of terms which should be included to obtain an accurate solution increases nonlinearly. Calculations show that at an orbital eccentricity of $0.95$, more than 700 terms of $G^2_{31q}$ (see Appendix A for the definitions of symbols used throughout the paper) are greater than 0.01 times the maximum value. 

Therefore, because the first approach of a planet to a white dwarf must be on an orbit with an eccentricity which exceeds 0.99, we cannot yet model the initial damping phase. Also, the planetary rotation becomes chaotic and unpredictable after the first passage close to the star, which makes the tidal evolution scenarios essentially probabilistic. Although eccentricities of about 0.8 can be feasibly modelled for individual systems, here we wish to perform an ensemble of simulations over a broad region of parameter space and to resemble the low-to-moderate eccentricities of the known planetesimals orbiting WD 1145+017 \citep{vanetal2015} and SDSS J1228+1040 \citep{manetal2019}. Therefore, we consider orbital eccentricities in the range of $10^{-4}$-$0.4$. In contrast to high eccentricities, high inclinations do not impose computational restrictions.

\subsection{Plan for paper}

In Section 2 and Appendices A--C, we set up and describe our planet-star
setup, the equations of motion, and the meaning of the various components of those equations.
Appendix A in particular is composed of two tables which describe all of the variables
used in this paper.
In Section 3, we identify results which can be gleaned through an inspection of the equations alone.
Section 4 is devoted to elucidating our initial conditions and parameters for the simulations.
In Section 5, we run the simulations and report the results. We conclude in Section 6.

\section{Setup}

In order to carry out our orbital simulations, we rely on the formalism presented
in \cite{bouefr2019}. That investigation corrects earlier work
and provides equations for the evolution of semimajor axis,
eccentricity and inclination of the mutual orbit of two bodies due to tides raised on both. The
formalism in that paper is general enough so that (i) either or both bodies may be stars or planets, and (ii)
the expansions do not diverge for high values of eccentricity nor inclination, unlike for some classical
expansions of planetary disturbing functions for the three-body problem \citep{murder1999}.

Our consideration will be limited to terrestrial planets
(like exo-asteroids and exo-Earths) and ice giants (like exo-Neptunes) 
as long as they do not have any surface continents nor surface oceans. 
Internal oceans are allowed because tidal dissipation in internal 
oceans is much lower than in mantles
\citep[Table 3 of][]{cheetal2014}. Surface oceans and continents
are not allowed because tidal
dissipation in such planets is dominated by friction in shallow
seas and cannot be described by rheological models borrowed
from continuum mechanics. Our method is not applicable to
gas giants (like exo-Jupiters); for descriptions of tidal friction 
on such objects, see the review by \cite{mathis2018} and references therein.

 \subsection{Orbital evolution owing to tides}

 We consider a planet orbiting a white dwarf.
 %  Here we assume that the primary body is a white dwarf and that the secondary body is a planet.
 Physical variables with a $\star$ subscript or superscript refer to a stellar physical property, and those without a superscript or subscript refer to a planetary physical property. As usual, the notations $a$ and $e$ refer to the semimajor axis and eccentricity of the mutual orbit. For inclination, $i$ refers to the orbit's inclination with respect to the planetary equator, while $i^{\prime}$ denotes the orbit's inclination with respect to the stellar equator. With this notation, the rates of $a$, $e$, $i$, and $i^{\,\prime}$ are given by equations (\ref{aevol}-\ref{ipevol}) in Appendix C. We do not consider the time evolution of the arguments of pericentre, longitudes of ascending node, and times of pericentre passages; the Lagrange-type orbital equations describing their rates are written down in \cite{bouefr2019}.

 The four orbital equations (\ref{aevol}--\ref{ipevol}) are secular: they are not explicit functions of the mean anomaly, true anomaly, nor mean longitude. This secular property allows us to model the system for longer times than if we considered the oscillatory and libratory behaviour on a per-orbit basis. We have used the bra-ket delimiters to indicate the secular nature of these equations. Also, these four evolution rates all receive double sets of delimiters because averaging is carried out not only over one orbit but also over one cycle of the apsidal precession.

 The expansions for each of the four orbital rates contain the inclination functions $\,F_{lmp}(i)\,$ and the eccentricity functions $\,G_{lpq}(e)\,$, which are given in equations (\ref{eccfunc}) and (\ref{incfunc}). These functions are not dependent on any physical parameters, and contain only a single orbital parameter each. Therefore, we precompute these functions before running any simulations. The values of $\,F_{lmp}(i)\,$ are straightforwardly computed for any inclination: the finite limits of the summation in equation (\ref{incfunc}) allow us to produce compact explicit formulae.

 However, convergence of the $\,G_{lpq}(e)\,$ values becomes computationally onerous for eccentricities near unity, because of the often infinite upper limit in one of the summations in equation (\ref{eccfunc}). To speed up computations, \cite{noyetal2014} tabulated values of $\,G_{lpq}(e)\,$ for different eccentricities and values of $l$, $p$, and $q$, and used the resulting look-up table in their simulations. Here, instead, for each given set of $l$, $p$, and $q$, we fit $G_{lpq}(e)$ to an explicit function of $e$. We then insert these explicit functions into the simulation arrays so that no lookup as a function of eccentricity is necessary.

 \subsection{Link to physical parameters\label{link}}

 In Appendix C, we write down the expressions (\ref{aevol}--\ref{ipevol}) for the tidal evolution rates of four orbital parameters: the mutual orbit's semimajor axis $a$, eccentricity $e$, inclination $i$ with respect to the planetary equator, and inclination $i^{\,\prime}$ with respect to the stellar equator. Each of those expressions is an infinite sum~---~over some integers $\,l,\,m,\,p,\,q\,$~---~of terms proportional to the so-called ``quality functions''. A quality function of degree $l$ equals the product of the degree-$l$ dynamical Love number and the sine of the degree-$l$ phase lag, both understood as functions of an $lmpq$ tidal Fourier mode.

 Specifically, the quality functions of the planet, $k_{l}(\omega_{lmpq}) \sin\epsilon_{l}(\omega_{lmpq})$, are functions of the tidal Fourier modes\footnote{~Be mindful that the functional form of a quality function is parametrised with the degree $\,l\,$ only, while the dependence on the other three integers comes through the arguments $\omega_{lmpq}$ or $\omega_{lmpq}^{\star}$. This simplification is available only if we approximate the body with a homogeneous sphere. The general case is more complex. For example, if we take into account the permanent oblateness of a body, its quality functions will be parametrised not only with the degree $l$, but also with the order $m$, and will read as $k_{lm}(\omega_{lmpq})\sin\epsilon_{lm}(\omega_{lmpq})\,$ and  $k_{lm}^{\star}(\omega^{\star}_{lmpq})\sin\epsilon^{\star}_{lm}(\omega^{\star}_{lmpq})$.
 In this paper, however, we will not delve into this level of complexity.}
  \begin{equation}
  \omega_{lmpq}\,\approx\,\left(l - 2p + q\right)n - m \frac{d\theta}{dt},
  \label{omegaplan}
  \end{equation}
 which are exerted on the planet by the star. Their absolute values
  are the actual forcing frequencies of deformation in the planet:~\footnote{~The physical meaning of the modes $\,\omega_{lmpq}\,$ and frequencies $\,\chi_{lmpq}\,$ can be easily understood through Eq. 15 in \cite{efrmak2013}.}
  \begin{equation}
 \chi_{lmpq} = \left|\omega_{lmpq}\right|.
 \label{chiplan}
 \end{equation}

 Similarly, the quality functions of the star, $k_{l}^{\star}(\omega_{lmpq}^{\star}) \sin\epsilon_{l}^{\star}(\omega_{lmpq}^{\star})$, are functions of the modes
  \begin{equation}
  \omega_{lmpq}^{\star} \approx \left(l - 2p + q\right)n - m \frac{d\theta^{\star}}{dt}
  \label{omegastar}
  \end{equation}
 which are excited inside the star by the planet. The absolute values of these modes,
 \begin{equation}
 \chi_{lmpq}^{\star} = \left|\omega_{lmpq}^{\star}\right|
 \label{chistar}
 \end{equation}
 are the physical frequencies of deformation inside the star.

 For both the planet and the star, the shapes of the quality functions are determined by interplay of self-gravitation and rheology in the appropriate body. Therefore, for each of the two bodies, the shape of an $l^{\rm th}$ quality function depends on this body's key kinematic and physical parameters, such as the spin rate, density, radius, surface gravity, shear viscosity, and shear compliance (which is the reciprocal of shear rigidity, or shear elastic modulus).

 Assume for simplicity that the tidally-perturbed celestial body is homogeneous and near-spherical, and also that its rheology is linear (which is usually the case). A general expression for a degree-$l\,$ quality function of such a body was derived in \citet[Eq. 169]{efroimsky2012a} and \citet[Eq. 40a]{efroimsky2015}. That expression is valid for an essentially arbitrary linear rheology. Mathematically, that expression is a simple algebraic functional of two functions representing the real and imaginary parts of the complex compliance of the material of which the celestial body is composed. Recall that these parts of the complex compliance contain the entire information about the (linear) rheology.

 Under the additional assumption that the planet's material adheres to the Maxwell model,\footnote{~This assumption restricts our planets to those where most of the tidal dissipation is taking place in the rocky material. The tidal response of terrestrial mantles is viscoelastic (Maxwell) at sufficiently low frequencies (for the Earth, up to about the Chandler frequency). At higher frequencies, the behaviour of these mantles is more adequately described by the Andrade model \citep{efroimsky2012a,efroimsky2012b} or, even better, by its generalisation named the Sundberg-Cooper model \citep{RenaudHenning2018}. It is very important that both these models are firmly rooted in physics, and reflect actual physical mechanisms of friction emerging at seismic frequencies. It is also convenient that, mathematically, they both are extensions of the Maxwell model.
Because the current paper is our pilot tidal publication on the topic of planets around white dwarfs, we take the liberty of omitting some technicalities with the understanding that these may be built in the theory later. Among these omitted items is the switch to more complex rheologies which is to be left for future work.}
 \,the quality functions of the planet become Eqs. 31 and 50a from
 \cite{efroimsky2015}:
 \[
 k_{l}(\omega_{lmpq})\;\sin\epsilon_{l}(\omega_{lmpq})
 \]
%\[
%=  k_{l}(\chi_{lmpq}) \sin{\left[\epsilon_{l}(\chi_{lmpq})\right]}
%  {\rm Sgn}\left(\omega_{lmpq}\right)
%\]
 \begin{equation}
 \ \ \  =\;
  \frac{3 \ {\rm Sgn}\left(\omega_{lmpq}\right)}{2\left(l-1\right)}\;
  \frac{  \frac{\textstyle\mathcal{B}_{l}}{\textstyle\eta\,\chi_{lmpq}}}{\left(\mathcal{J}
           + \mathcal{B}_{l}\right)^2 + \left(\eta\chi_{lmpq}\right)^{-2} }
 \label{qualplan}
 \end{equation}
 \noindent{}where
 \begin{equation}
  \mathcal{B}_{l} = \frac{3 \left(2l^2 + 4l + 3\right)}{4 l \pi \mathcal{G} \rho^2 R^2}
          = \frac{2l^2 + 4l + 3}{l g \rho R}.
\label{planB}
\end{equation}
  \noindent{}Here $\cal G\,$ is Newton's gravitational constant, while $\rho$, $R$, $g$, $\eta$, and ${\mathcal{J}}$ are the planet's mean density, radius, surface gravity, shear viscosity, and shear compliance, correspondingly.

 Under the assumption that the star, too, has a Maxwell rheology,\,\footnote{~Stars are commonly assumed to be viscous \citep{stix2002}. Simultaneously, there exist theoretical indications that they may possess magnetic rigidity \citep{williams2004,williams2005,williams2006,ogilvie2008,garaudetal2010}. We are hence motivated to treat a stellar material, at large, as a Maxwell body, though with a very small shear compliance $\cal{J}^{\star}$. Not knowing {\it{how small}} the compliance value may be, we assume that  $\cal{J}^{\star}$ is much smaller than $1/(\eta^{\star}\chi^{\star})$, where $\eta^{\star}$ and $\chi^{\star}$ are the mean viscosity of the star and a typical tidal frequency, correspondingly. This approximation simplifies the expression for the stellar quality functions \citep[Eq. 60]{efroimsky2015}.} \,the quality functions of a star approximated with a homogeneous Maxwell sphere are
 \[
  k_{l}^{\star}(\omega_{lmpq}^{\star})\;\sin\epsilon_{l}^{\star}(\omega_{lmpq}^{\star})
 \]
%\[
%= k_{l}^{\star}(\chi_{lmpq}^{\star}) \sin{\left[\epsilon_{l}^{\star}(\chi_{lmpq}^{\star})\right]}
%  {\rm Sgn}\left(\omega_{lmpq}^{\star}\right)
%\]
 \begin{equation}
 \ \ \ =\;
 \frac{3 \ {\rm Sgn}\left(\omega_{lmpq}^{\star}\right)}{2\left(l-1\right)}\;
 \frac{  \frac{\textstyle\mathcal{B}_{l}^{\star}}{\textstyle\eta^{\star}\,\chi_{lmpq}^{\star}}}{\left(\mathcal{J}^{\star}
           + \mathcal{B}_{l}^{\star}\right)^2 + \left(\eta^{\star}\,\chi_{lmpq}^{\star}\right)^{-2} }
 \label{qualstar}
 \end{equation}
 \noindent{}where
 \begin{equation}
 \mathcal{B}_{l}^{\star} = \frac{3 \left(2l^2 + 4l + 3\right)}{4 l \pi \mathcal{G} \rho_{\star}^2 R_{\star}^2}
                         = \frac{2l^2 + 4l + 3}{l g_{\star} \rho_{\star} R_{\star}},
 \label{starB}
 \end{equation}
 \noindent{}with $\rho^{\star}$, $R^{\star}$, $g^{\star}$, $\eta^{\star}$ and ${\mathcal{J}}^{\star}$ being the stellar mean density, radius, surface gravity, shear viscosity, and shear compliance, accordingly.

 For a comprehensive list of variable definitions, see Tables A1--A2.

\subsection{Spin equations of motion}

The orbital evolution equations of motion (\ref{aevol}--\ref{ipevol}) contain explicit dependences on the spin evolution of the planet and star ($d \theta / dt$ and $d \theta^{\star} / dt$) through equations (\ref{omegaplan}--\ref{omegastar}). Therefore spin evolution must be solved self-consistently with orbital evolution. We provide the general spin evolution equations for both the star and planet in equations (\ref{thetaevol}--\ref{thetaevolstar}).

These equations represent the sum of the contributions from both the triaxial torque and tidal torque. The secular triaxial torque (as opposed to the non-secular version) is conventionally accepted to equal zero (equations \ref{tri}--\ref{tristar}) because the averaging (i) would otherwise be certain to yield a non-zero secular term only in physically questionable circumstances (synchronicity and nonzero net tilt) and (ii) might yield a non-zero secular term in other largely unexplored circumstances (e.g. due to the variation of the initial mean anomaly, argument of pericentre, and $\dot{\theta}$ from oblateness terms). The secular tidal torque is given by equations (\ref{tid1}--\ref{tid1star}).

\section{Properties of the equations of motion}

Some results may be gleaned just through inspection of equations (\ref{aevol})--(\ref{thetaevolstar}), without
needing to run any numerical simluations.

\subsection{Synchronisation}

The spin and orbital element evolution dependencies on multiple variables suggest that a simple
characterisation of spin-orbit resonances is difficult to attain. A direct comparison
of the spin periods of the white dwarf and planet, along with their mutual separation,
would alone be insufficient to claim  synchronicity or not. Resonant trapping of an individual
system is best devoted to a dedicated study (e.g. \citealt*{noyetal2014} for Mercury
or \citealt*{maketal2012} for GJ 581d). Further, the secular (averaged) nature of the equations applied here
might miss important librational behaviours which occur on orbital timescales, and can play
a role in resonant capture.

\subsection{Circularisation}

Strictly, a perfectly circular orbit will remain circular in the midst of tidal interactions,
as can be demonstrated formally by taking the limit of equation (\ref{eevol}) as $e \rightarrow 0$.
However, realistically, this situation does not occur: even a minuscule occasional interaction
capable of generating a nonzero value of $e$, will, under some circumstances, initiate a secular
growth in $e$. Therefore, equation (\ref{eevol}) should be integrated simultaneously with the other
orbital and spin equations regardless of how small $e$ is thought to become.

\subsection{Planarisation}

When the equators of both the planet and the white dwarf are perfectly aligned with the orbit,
then this situation will remain unchanged in the midst of tidal interactions.
Importantly, however, in this case (when $i = i' = 0$), the inclination $F_{lmp}$ functions do
not vanish for all $l,m$ and $p$. Nevertheless, the coplanar case does help reduce computational cost
because in that case, only those sets of $l,m$ and $p$ for which $F_{lmp}$ does not vanish need to be
included in the summations in equations (\ref{aevol}--\ref{eevol}).

\subsection{Contributions from stellar quality functions}

 The semimajor axis and eccentricity rates (\ref{aevol}--\ref{eevol}) feature distinct terms associated with each quality function.
 Orbital evolution is then said to be driven by planetary tides when the planetary quality function dominates, and driven by stellar tides when the stellar quality function dominates.   
When both tidally interacting bodies are of similar types (such as in a binary star system or a binary asteroid system),
 then a degree-$l\,$ term owing to the tides in one body and an analogous degree-$l\,$ terms owing to the tides in another body may make comparable contributions to the overall evolution.

However, for star-planet systems, the stellar quality function coefficients $M/M_{\star}$ and $(R_{\star}/a)^{2l+1}$ are small. The second term becomes even smaller for white dwarfs, because their radii are comparable to the Earth's radius rather than the Sun's radius. The smallness of these coefficients often~---~but not always~---~makes negligible the stellar quality function terms. Exceptions include when the stellar quality function itself is large, which occurs when the stellar viscosity is large and/or when the star spins quickly. One potential manifestation of this exception is an eccentricity increase when the stellar spin is faster than $(3/2)n$ \citep{bouefr2019}.

For the numerical investigations in this paper, the stellar quality function remains small~---~primarily because of our adopted white dwarf viscosity~---~greatly helping to facilitate the reduction of the phase space that we explore. Hence, because the equations for spin evolution (\ref{thetaevol}--\ref{thetaevolstar}) are each a function of just one of the quality functions, we find that the white dwarf spin rate changes negligibly due to tidal effects with a planet, and can be considered fixed. Similarly, each term in the evolution of $i^{\,\prime}$ (equation \ref{ipevol}) is a function of the white dwarf quality function, but not the planetary quality function. Hence $i^{\,\prime}$ may also be considered fixed. The integrations of the full equations of motion in Section 5 confirm the resistance of $d\theta^{\star}/dt$ and $i^{\,\prime}$ to change.

 \subsection{Asteroidal tides}

 The expressions for the orbital elements' rates from \cite{bouefr2019} apply to any two near-spherical bodies, asteroids included.  As mentioned in Section 1, asteroids play a key role in white dwarf planetary systems. The asteroid (or asteroids) discovered disintegrating around WD 1145+017 \citep{vanetal2015} has now been the subject of over 20 dedicated refereed papers
\citep{aloetal2016,gaeetal2016,rapetal2016,xuetal2016,zhoetal2016,croetal2017,faretal2017,garetal2017,guretal2017,haletal2017,kjuetal2017,veretal2017, cauetal2018,faretal2018,izqetal2018,rapetal2018,xuetal2018,duvetal2019,karetal2019}. Further, the asteroid embedded well-within the Roche radius (for rocky compositions) of the white dwarf SDSS J1228+1040 \citep{manetal2019} resides in one of the most dynamically active white dwarf planetary systems.

 Nevertheless, an outstanding question remains about the origin and dynamical pathways of these asteroids. If an asteroid were originally spherical, could it have been tidally torqued into the white dwarf Roche radius? The equations here cast doubt on this scenario. The effect of tides become weaker as the planetary radius decreases:  for a constant density, the planetary quality function
 $\,\sim 1/\mathcal{B}_l \sim g \rho R \sim M/R\,$, and its coefficients are $\,\sim R^{2l+1}/M\,$, giving
 \begin{equation}
 \left\langle \left\langle
 \frac{da}{dt}
 \right\rangle \right\rangle
 \sim R^{2l},
 \label{steep}
 \end{equation}

 \begin{equation}
 \left\langle \left\langle
 \frac{de}{dt}
 \right\rangle \right\rangle
 \sim R^{2l},
 \end{equation}

 \begin{equation}
 \left\langle \left\langle
 \frac{di}{dt}
 \right\rangle \right\rangle
 \sim R^{2l},
 \end{equation}

\noindent{}where $l \ge 2$.
The steep dependence of these rates on the planetary radius illustrates that the tides in near-spherical asteroids are negligible
and could not push the asteroids into the white dwarf Roche radius. Other mechanisms must be invoked (see Section 1.1).
 %  In reality, however, asteroids are likely to be aspherical, and the resulting permanent-triaxiality-caused
 %  torque (equations \ref{tri} - \ref{tristar}) could render quite a different result.

\section{Initial conditions and parameters for simulations}

An important feature of this paper is not only the establishment of the
tidal equations, but also the determination what physical situations to integrate.
Our simulations require 16 initial conditions and parameters to be established:

\begin{itemize}

\setlength\itemsep{0.7em}

\item 4 for the orbit: $a(0), e(0), i(0), i'(0)$

\item 6 for the planet: $\theta(0)$, $\frac{\textstyle d\theta}{\textstyle dt}(0)$, $M$, $R$, $\mathcal{J}, \eta$

\item 6 for the star: $\theta^{\star}(0)$, $\frac{\textstyle d\theta^{\star}}{\textstyle dt\,}(0)$, $M_{\star}$, $R_{\star}$,
$\mathcal{J}^{\star}, \eta^{\star}$

\end{itemize}

\noindent{}The masses, radii, shear viscosities, and shear compliances  \footnote{~We remind the reader that the shear compliance is the inverse
of the shear elastic modulus.} (equalling eight variables) are all assumed to remain fixed in time throughout
each individual simulation.  Hence we ignore the potential change in planetary viscosity as the planet approaches the white
dwarf, a phenomenon which has been speculated to occur in the TRAPPIST-1 system \citep{maketal2018}. We also 
ignore the plausible possibility that a planet's mass and radius will change
due to, for example, sublimation. Such sublimation, however, has been shown to have a negligible effect on the orbital pericentre
\citep{veretal2015c} but could have a larger effect through the reduction of planetary mass. Stellar evolution is irrelevant
for white dwarfs on the maximum timescale of our simulations (100 Myr).

Both the star and planet are assumed to be represented fully by the 12 physical parameters
above. For example, a single value of $\eta$ must be applied for the entire planet (as opposed to, for example, separate values for its mantle and core).

We cannot, in this paper, cover the entire phase space encompassed by these parameters. Therefore, we must carefully choose which ensembles of variables to vary across our simulations, and do so with computational limitations in mind (see Appendix C).

\subsection{Fixed initial conditions and parameters across simulations}

In total, we choose to fix
10 variables ($\mathcal{J}$, $\mathcal{J}^{\star}$, $i(0), i'(0)$, $\theta(0), \theta^{\star}(0), (d \theta^{\star}/dt)(0)$,
$M_{\star}, R_{\star}, \eta^{\star}$) of the 16, for the following reasons:

\begin{itemize}

\item {\bf Compliances.} As explained in Footnote 4, a star may possess some effective magnetic rigidity.  We, however, assume that the inverse of rigidity -- the compliance -- is small ($\mathcal{J}^{\star} \ll 1 / (\eta^{\star} \chi^{\star})$), such that the elastic reaction of the stellar material to the tidal stress is much less important than its viscous reaction (see equation \ref{qualstar}). Therefore, we set $\mathcal{J}^{\star} = 0$~(1/Pa).

The real compliance $\mathcal{J}$ of planets, asteroids, and comets is a better-known quantity:
for the solid Earth, $1/{\mathcal{J}}~\approx~0.8~\times~10^{11}$~Pa; for ices, $1/{\mathcal{J}}~\approx~4~\times~10^9$~Pa; and for snow, $1/{\mathcal{J}}~\approx~10^6$~Pa. The corresponding values for ${\mathcal{B}}_l$ are such that, for small bodies and for terrestrial planets smaller than Earth, ${\mathcal{J}}~\ll~{\mathcal{B}}_l$, so that ${\mathcal{J}}$ plays no role in equation (\ref{qualplan}) (provided that the rheology is Maxwell). For overheated super-Earths (which are expected to obey the Maxwell model, see \citealt*{makarov2015}), we still can neglect ${\mathcal{J}}$, because both ${\mathcal{J}}$ and ${\mathcal{B}}_l$ are much smaller than $1/(\eta\chi_{lmpq})$ in the denominator of the right-hand side of equation (\ref{qualplan}). However, in the case of colder super-Earths, the term ${\mathcal{J}}$ in the denominator of that expression cannot be neglected.

\item {\bf Initial spin orientations.} Our numerical integration results are insensitive to $\theta(0)$ and $\theta^{\star}(0)$, at least for the outputs of interest. Therefore, we set  $\theta(0) = \theta^{\star}(0) = 0^{\circ}$.

\item {\bf Initial spin rate of white dwarf.} Because equation (\ref{thetaevol}) is a second-order differential equation, we must set the initial spin rate ($d\theta^{\star}/dt$)(0) of the white dwarf in addition to initial orientation $\theta^{\star}(0)$. As argued in Section 3.4, in most cases the white dwarf's spin rate will change negligibly due to tidal interactions with a planet, and will negligibly affect other values of interest through the white dwarf quality function. Nevertheless, we still must set a value.

White dwarf spin rates range from $0-1$ revolution/hour, where the upper bound corresponds to the shortest rotational periods found from the observed population of single pulsating white dwarfs \cite[Fig. 4 of][]{heretal2017}\footnote{These values give an equatorial rotational speed range of about 0-15 km/s.}. However, a more typical upper bound is likely to be 1 revolution/day (corresponding to an equatorial rotational speed of about 0.6 km/s), as 1 revolution/hour represents extreme cases where the white dwarf's spin was likely kicked due to a former binary companion. Hence, we adopt $(d\theta^{\star}/dt)(0) =$ 1 revolution per $20$ hours.

\item {\bf Initial orbital inclinations.} As argued in Section 3.4, the evolution of $i^{\prime}$ negligibly affects $a$ and $e$ and does not affect $i$ nor $d\theta/dt$ at all.  Hence, here we arbitrarily set $i'(0) = {10^{-4}}^{\circ}$.

 In our preliminary simulations, we found that the orbital inclination $i$ usually either (i) decreases to the coplanar limit sharply relative to the semimajor axis and eccentricity evolution, or (ii) increases slightly.  This distinction depends on the relative strength of the two terms in equation (\ref{ievol}), and was consistent across all of our preliminary simulations. Further, the magnitude of the inclination, even if above $90^{\circ}$, does not affect this bimodal outcome, and does not substantially affect the semimajor axis nor eccentricity evolution. Therefore, we do not learn enough by varying $i(0)$ to justify doing so. Hence, purely for demonstration purposes, we set $i(0) = 140^{\circ}$.\footnote{Although this retrograde value might seem high, planets have been shown to reach near the white dwarf Roche radius on arbitrarily highly inclined orbits in both single or binary systems \citep{vergae2015,hampor2016,petmun2017,steetal2017,veretal2018a}.}

\item {\bf White dwarf mass.} Conveniently, the mass distribution of the population of white dwarfs is highly peaked around $0.60-0.65 M_{\odot}$ \citep{treetal2016}, and here we adopt $M_{\star} = 0.60M_{\odot}$.

\item {\bf White dwarf radius} White dwarf radii are closely linked with their masses (see e.g. equations 27-28 of \citealt*{nauenberg1972}, equation 15 of \citealt*{verrap1988}, and \citealt*{bosque2018}). Consequently, given our choice of $M_{\star} = 0.60M_{\odot}$, we choose $R_{\star} = 0.01280 R_{\odot} = 8900$ km.

\item {\bf White dwarf viscosity} The star's dynamic shear viscosity is a parameter for which observational constraints have proven largely elusive. We employ a simple approximation and treat the entire star as a viscous sphere (with a constant viscosity). Even this simple approximation, however, cannot hide the gaping uncertainty in our knowledge of white dwarfs' viscosity. \cite{dalros2014} suggest that $\eta^{\star} \sim 10^7-10^{17}$ Pa$\cdot$s, depending on whether the main driver of the viscosity is radiation, plasma, magnetism, turbulence or a combination thereof. Despite this large range, we fix a specific value through the following procedure.

A method to reduce this uncertainty is to treat the shear viscosity as the product of the star's average density ($\rho_{\star}$),
the typical scale of convective flows ($d_{\star}$, which is comparable to the typical size of supergranules), and the typical speed
in the supergranules ($v_{\star}$). An extension of equation 7.32 of \cite{stix2002} gives

\begin{equation}
\eta^{\star} = \rho_{\star}d_{\star}v_{\star}.
\label{granule}
\end{equation}

\noindent{}For the Sun, $v_{\odot} \approx 300$ m/s, $d_{\odot} = 1.6 \times 10^7$ m and $\rho_{\odot}~=~200$~kg/m$^3$,
yielding $\eta_{\odot} = 1.0 \times 10^{12}$ Pa$\cdot$s. \footnote{~Although \cite{stix2002} states on page 267 that a typical cell diameter is $\,1.6 \times 10^4$ km, we assume that he meant radius because his ensuing estimate for the mean spacing between cell centres is $\,3 \times 10^4$ km.}

We perform a similar computation for a variety of potential hydrogen-atmosphere based white dwarf hosts (known as ``DA white dwarfs'') by using the stellar models from \cite{koester2009} and \cite{koester2010}. DA white dwarfs are the most common types of white dwarfs. These models output convective speeds ($v_{\star}$) and the density in the envelope, which is more appropriate to use than the overall white dwarf density in order to model the convection zone. The models also output pressure, which can be combined with a given surface gravity to obtain the pressure scale height (Eq. 1 of \citealt*{treetal2013}), which in turn can be used to represent $d_{\star}$.

In Table 1, we list dynamic viscosities for DA white dwarfs at a location in the convection zone where the total white dwarf mass is a factor of 10 orders of magnitude more than the convective zone mass above that layer. The effective temperature range in the table from 5000 K to 9000 K corresponds to cooling ages of about 1 Gyr~--~5 Gyr (``cooling age'' is the time since the white dwarf was born). The table illustrates overall that the dynamic viscosity of these white dwarfs are near the lower end of the range proposed by \cite{dalros2014}, but are nevertheless relatively well-confined. Hence, we fix $\eta^{\star} = 10^7$ Pa$\cdot$s.

\end{itemize}

\begin{table}
% \centering
% \begin{minipage}{180mm}
 \centering
  \caption{Self-consistent estimations of DA white dwarf dynamic (or shear) viscosities in the convective zone
(column 3), masses (column 4), and radii (column 5) for given values of effective temperatures (column 1),
and stellar luminosities (column 2). A surface gravity of $g_{\star} = 10^6$ m/s$^2$ was assumed.
Values were taken at a depth corresponding
to a convection zone mass above that layer which is ten orders of magnitude smaller than the white dwarf mass.
}
  \begin{tabular}{ccccc}
  \hline
$T_{\rm eff}^{\star}$/K & $L_{\star}$/$L_{\odot}$ & $\eta^{\star}$/(Pa$\cdot$s) & $M_{\star}$/$M_{\odot}$  &  $R_{\star}$/$R_{\odot}$   \\
 \hline
5000 &  $9 \times 10^{-4}$ &  $2.49 \times 10^6$   & 0.579  & 0.01260    \\
5500 &  $1.3 \times 10^{-3}$  &  $4.02 \times 10^6 $  & 0.583  & 0.01264  \\
6000 &  $1.9 \times 10^{-3}$  &  $5.46 \times 10^6$  & 0.588  & 0.01269  \\
6500 &  $2.6 \times 10^{-3}$  &  $6.85 \times 10^6$  & 0.590  & 0.01272  \\
7000 &  $3.5 \times 10^{-3}$  &  $8.23 \times 10^6$  & 0.592  & 0.01273  \\
7500 &  $4.7 \times 10^{-3}$  &  $9.38 \times 10^6$  & 0.593  & 0.01275  \\
8000 &  $6.1 \times 10^{-3}$  &  $9.95 \times 10^6$  & 0.596  & 0.01278  \\
8500 &  $7.7 \times 10^{-3}$  &  $1.20 \times 10^7$  & 0.596  & 0.01278  \\
9000 &  $9.8 \times 10^{-3}$  &  $1.33 \times 10^7$  & 0.598  & 0.01280  \\
 \hline
\end{tabular}
%\end{minipage}
\end{table}

%%%%%%%%%%%%%%%% Figure
\begin{figure*}
\centerline{
\includegraphics[width=8cm]{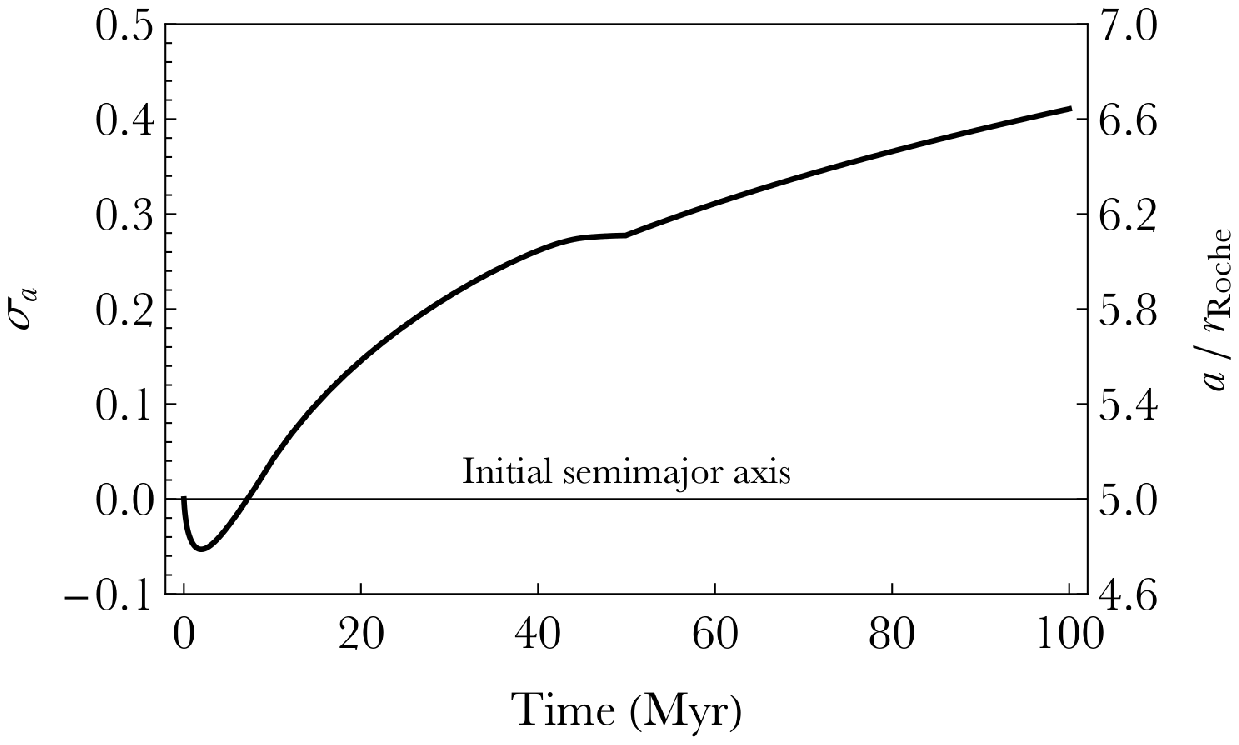}
\ \ \ \ \ \
\includegraphics[width=8cm]{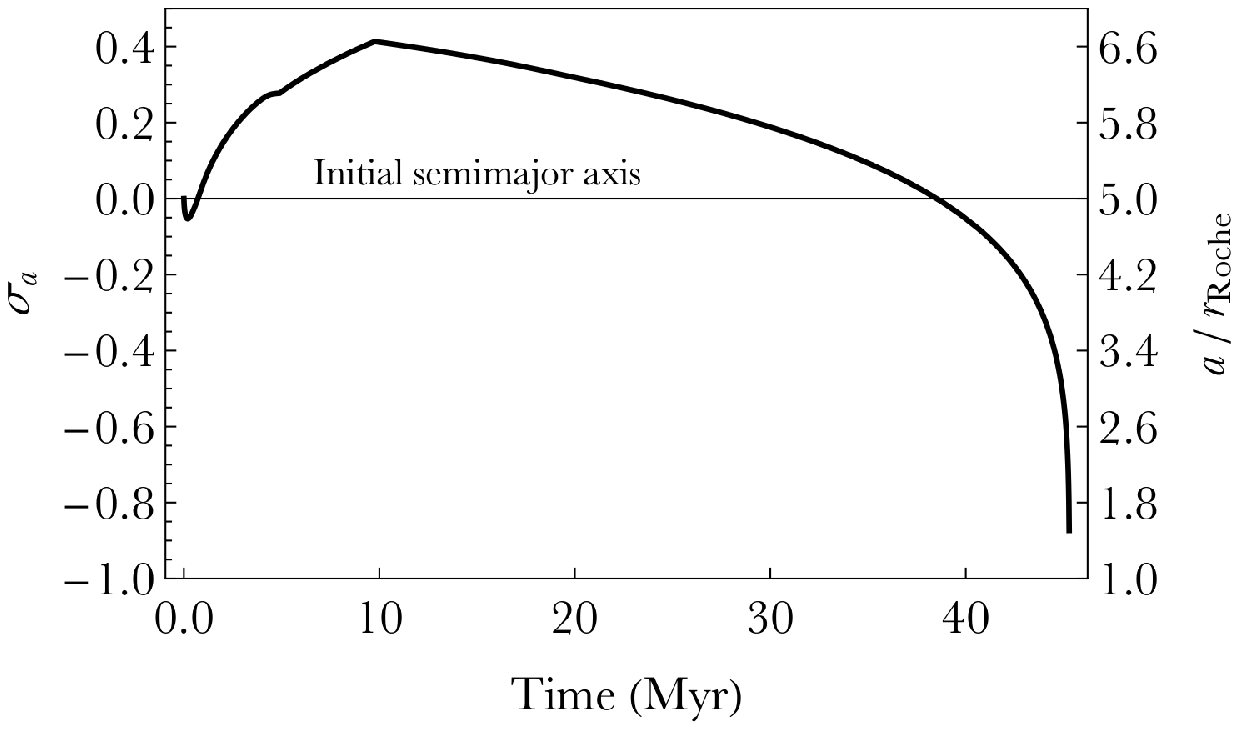}
}
\centerline{
\includegraphics[width=8cm]{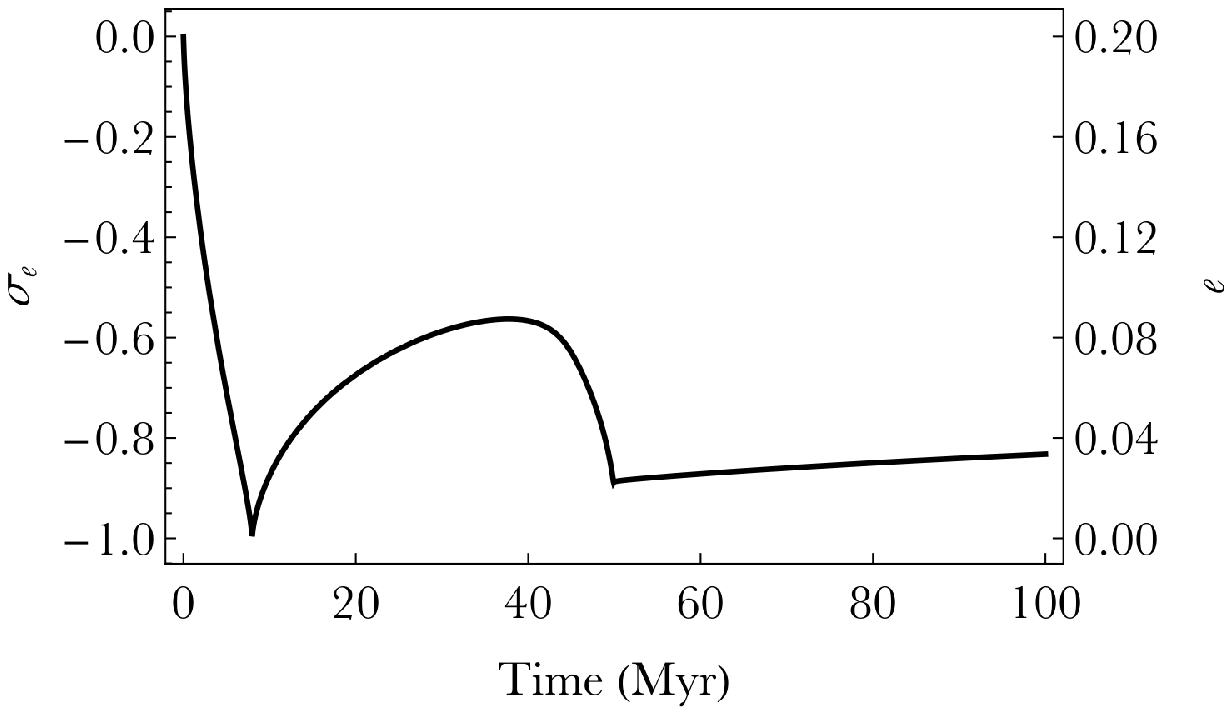}
\ \ \ \ \ \
\includegraphics[width=8cm]{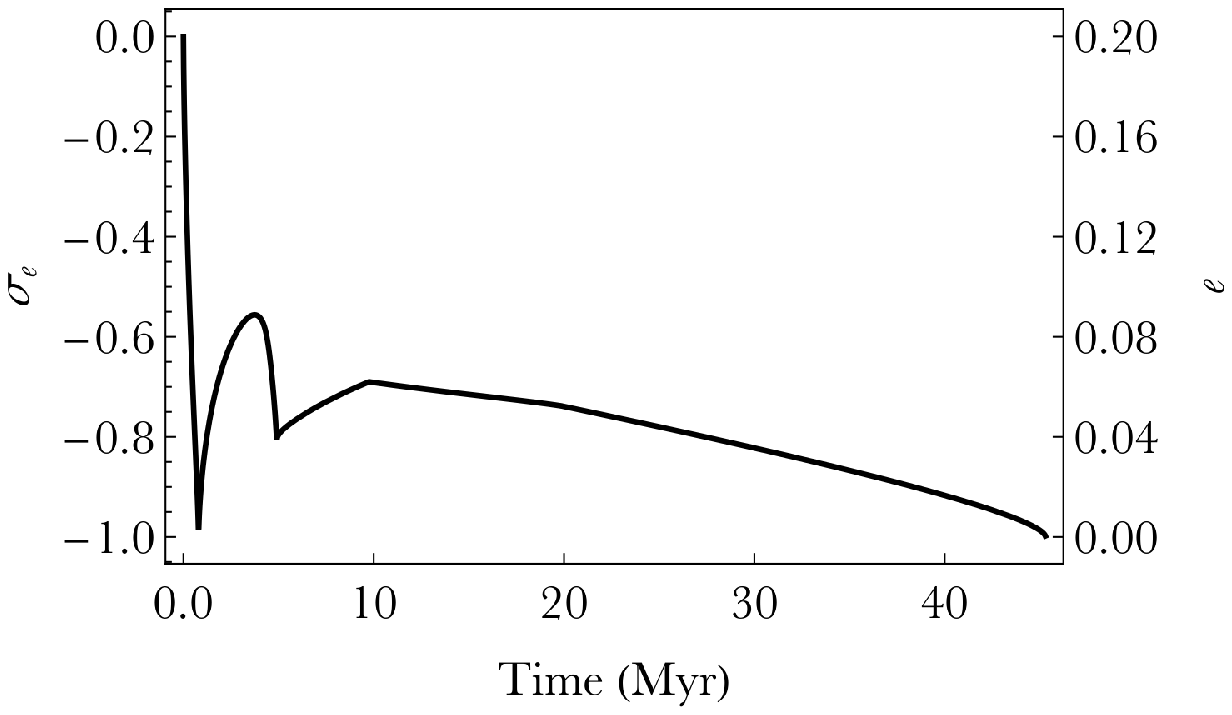}
}
\centerline{
\includegraphics[width=8cm]{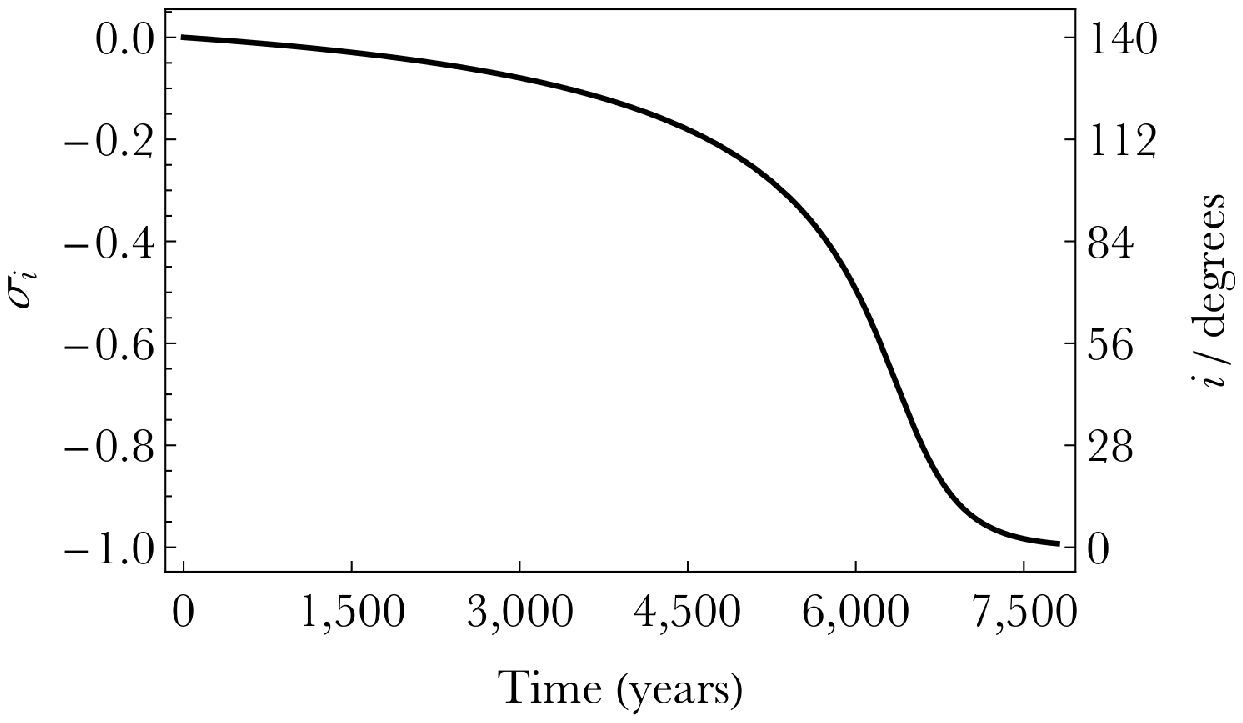}
\ \ \ \ \ \
\includegraphics[width=8cm]{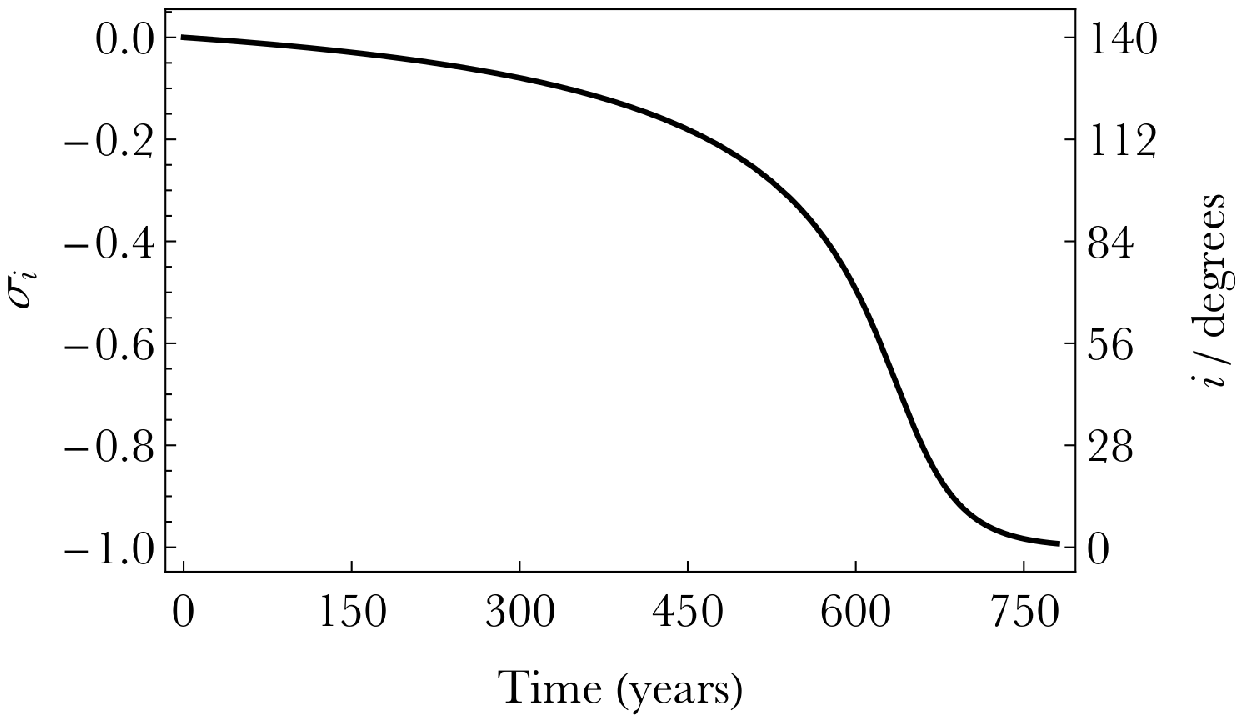}
}
\caption{
Two examples (left panels and right panels) of the orbital evolution of a planet tidally interacting
with a white dwarf, where definitions of $\sigma_a$, $\sigma_e$ and $\sigma_i$ are given in
equations (\ref{sigmaa}), (\ref{sigmae}) and (\ref{sigmai}). In both examples, $a(0) = 5r_{\rm Roche}$,
$e(0) = 0.20$,  $M = 10M_{\oplus}$, $\rho = \rho_{\oplus}$, and $(d\theta/dt)(0) = 360^{\circ}/ (20 \ {\rm hours})$. In the left panels, $\eta = 10^{20}$ Pa$\cdot$s, whereas in the right panels,
$\eta = 10^{19}$ Pa$\cdot$s; i.e. the only difference in the simulations is a one order of magnitude variation in the planetary viscosity. In the left panels, over 100 Myr, the semimajor axis evolution is increased overall, but in a non-uniform manner and only after shifting direction. In contrast, in the right panels, the planet changes direction twice, and is eventually destroyed by entering the Roche radius after 54 Myr. In both panels, the initial inclination of
the orbit $i(0) = 140^{\circ}$ with respect to the planet's equator is quickly reduced to within $1^{\circ}$. The damping of $i'$ (inclination with respect to the stellar equator) is even quicker (note the different scales of the time axes).
}
\label{cases}
\end{figure*}
%%%%%%%%%%%%%%%% Figure

\subsection{Variable initial conditions and parameters across simulations}

We vary only 6 variables ($a(0), e(0), M, R, (d \theta/dt)(0), \eta$) amongst the different simulations.

\begin{itemize}

\item {\bf Initial semimajor axis and eccentricity.}  Planets are assumed to approach the white dwarf in the
first instance on a highly eccentric ($>0.99$) orbit. Although the Darwin-Kaula expansion can
model any eccentricity under unity, practically the number of terms which
would need to be retained to compute an accurate solution for $e=0.99$ renders such simulations computationally infeasible.
Details of the initial damping period hence remain unknown for now, and would be enlightened in a future study
that could perhaps reformulate the Darwin-Kaula formalism by building an expansion over $e$ not about the value $e=0$, but about $e=1$.

Therefore, here we consider the planet at a later stage, after its eccentricity and semimajor axis have already been
moderately damped. Of interest to us is the rate of the subsequent inward or outward drift and whether or not the
planet will reach the white dwarf's Roche radius $r_{\rm Roche}$.
%up to a Hubble time (about 13 Gyr).
If we assume the planet is solid and spinning, then we could use the relevant Roche radius coefficient from \cite{veretal2017},
and with $M_{\star} = 0.60M_{\odot}$, we then obtain

\begin{equation}
\frac{r_{\rm Roche}}{R_{\odot}} \approx \left( \frac{\rho}{3 \ {\rm g \ cm}^{-3}} \right)^{-1/3}
.
\end{equation}

By using the Roche radius as a scaling, we adopt  $a(0)$ values ranging
from $2r_{\rm Roche}$ to $30r_{\rm Roche}$, with $10^{-4}~\le~e(0)~\le~0.4$. As explained in Section 3.2,
the eccentricity is never realistically exactly zero.

\item {\bf Planet mass.} One advantage of utilizing the Darwin-Kaula expansion of \cite{bouefr2019} is that the primary and secondary
could be any objects, including ones of asteroid-size (which is particularly relevant for white dwarf planetary
systems). Regardless, as explained in Section 3.5, orbital evolution of spherical asteroids is unaffected by tidal torques. 
Hence, we explore a selection of higher rocky masses. We vary the planet mass ($M$) from  $10^{1} M_{\oplus}$ 
(representative of a ``Super-Earth'')
to $10^{-3} M_{\oplus}$ (about the mass of Haumea or Titania). The trends in the results become obvious enough that
reducing the lower limit further is not worth spending the computational resources.

\item {\bf Planet radius.} We then obtain $R$ by assuming an appropriate value of $\rho$. Here, we simply adopt Earth's density for all values of $M$.

\item  {\bf Initial spin rate of planet.} As evidenced by the major planets in our solar system, planetary spin periods can
range from about 10 hours to $-$243 days (with the minus indicating retrograde motion); extrasolar planets may have a larger range,
but their spins are currently not as well constrained. The direction of planetary spin may alter the results, and hence we take that
case into account.  Here, we
adopt a slightly smaller range to what is seen in the solar system, with $(d\theta/dt)(0) =$ 1 revolution per $10-10^3$ hours,
in both the prograde and retrograde senses with respect to the orbital motion around the star.

\item {\bf Viscosity of planet.} The final parameter to specify, $\eta$, can vary significantly depending on the planet type. One recent
estimate of stagnant lid planets \citep{thietal2019} sets $\eta~=~10^{20}~-~10^{21}$~Pa$\cdot$s for dry mantle rheologies.
On Earth, for the Tibetan crust and lithosphere, $\eta~=~10^{18}~-~10^{20}$~Pa$\cdot$s \citep{henetal2019}, whereas for the 
Earth's mantle, $\eta~=~10^{20}~-~10^{24}$~Pa$\cdot$s \citep{mitfor2004}.

Several recent papers on the icy satellites in the solar system specify a wide range of dynamic viscosities.
Table 2 of \cite{huretal2018} lists $\eta = 10^{21}, 10^{21},10^{20}$ and $10^{14}$ Pa$\cdot$s respectively
for Europa's iron core,
brittle ice layer, silicate mantle, and ductile ice layer. \cite{cametal2019} give similar values
for Ganymede, whereas Table 1 of \cite{patetal2019} lists $\eta = 10^{28}, 10^{22}, 10^{14}$ Pa$\cdot$s
respectively for the core, upper ice layer and lower ice layer of Enceladus. The latter two values
actually were chosen from wide ranges previously reported in the literature of
$10^{19} - 10^{26}$ Pa$\cdot$s for the upper ice layer and $10^{12} - 10^{17}$ Pa$\cdot$s
for the lower ice layer.

Overall, the values given in the last paragraph indicate that adopting a range of $\eta = 10^{16} - 10^{24}$ Pa$\cdot$s is representative of most cases, and we do so.

\end{itemize}

%Because our treatment here is secular, modelling escape and capture into spin-orbit resonances would not be reliable (and would necessitate utilising unaveraged equations).

%The link established between orbital elements and physical parameters from equations (\ref{qualstar})-(\ref{planB}) assumes here that $\eta$ and $\eta^{\star}$ are represented by a single value at each timestep.

%%%%%%%%%%%%%%%% Figure
\begin{figure*}
\includegraphics[width=15cm]{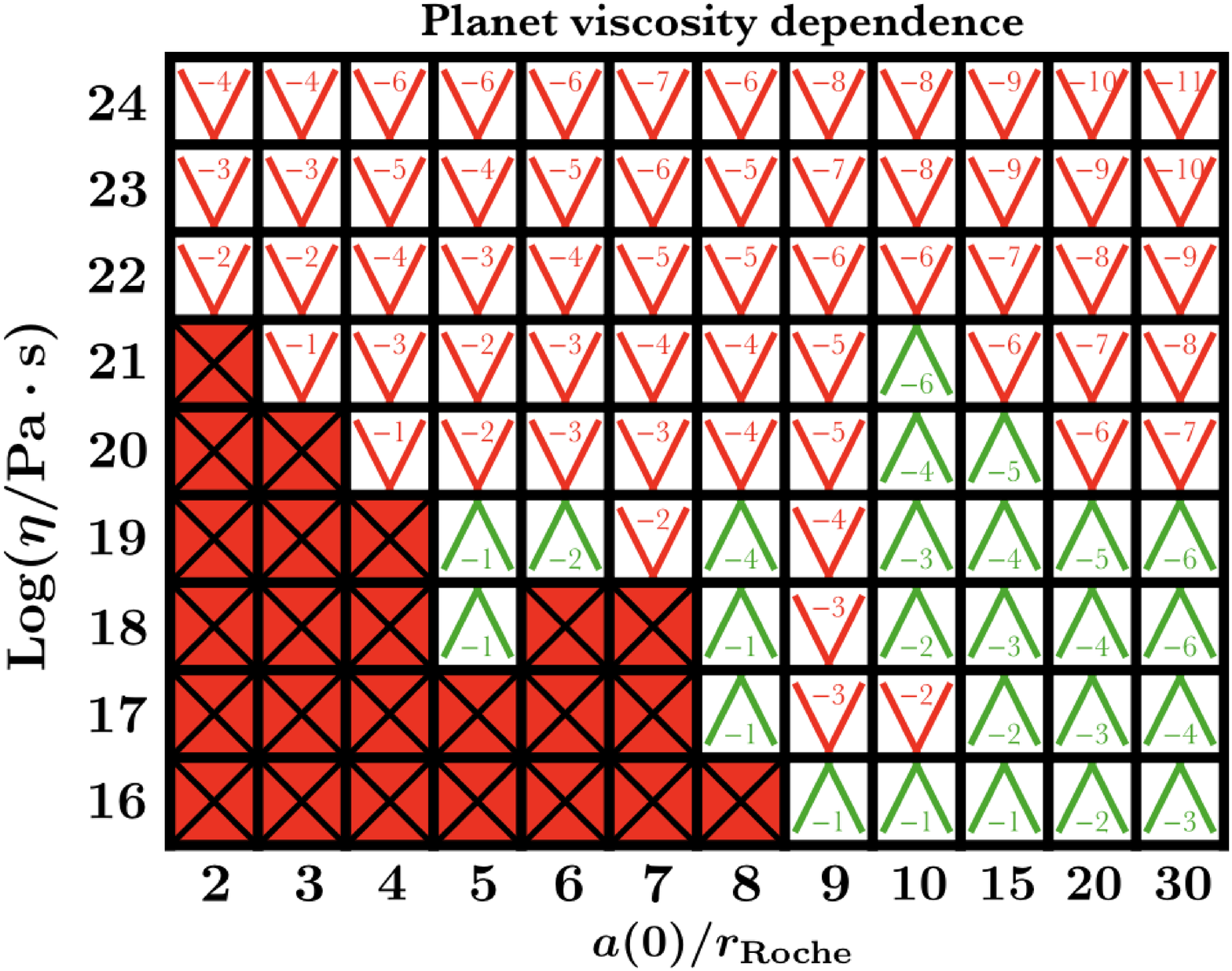}
\caption{
Outcomes of planetary movement due to tidal interactions with a white dwarf. For our adopted rheological model, a single value for planetary dynamic viscosity ($y$-axis) does not easily map to a specific known rocky and differentiated body, although a relatively homogenous dirty snowball like Enceladus corresponds well with $\eta \approx 0.24 \times 10^{14}$ Pa$\cdot$s \citep{efroimsky2018}.
Each box represents the state of a single simulation after 100 Myr of evolution.
Red crosses indicate that the planet's semimajor axis
has come within the white dwarf's Roche radius $r_{\rm Roche}$, destroying the planet. Otherwise,
the net migration of the planet after 100 Myr is either outward (green carets) or inward (red ``V''s). The number
within those symbols gives the magnitude of the migration, and is of the order of log$|\sigma_a|$. Other variables
which were assumed for these simulations are $e(0) = 0.2$, $M = M_{\oplus}$, $R = R_{\oplus}$, and
$(d\theta/dt)(0) = 360^{\circ}/ (20 \ {\rm hours})$. This figure illustrates that the outcome is strongly dependent on planetary viscosity, and suggests the existence of a fractal boundary and chaos due to the largest outward migrations neighbouring destructive spiral-ins.
}
\label{etaplot}
\end{figure*}
%%%%%%%%%%%%%%%% Figure

%%%%%%%%%%%%%%%% Figure
\begin{figure*}
\includegraphics[width=15cm]{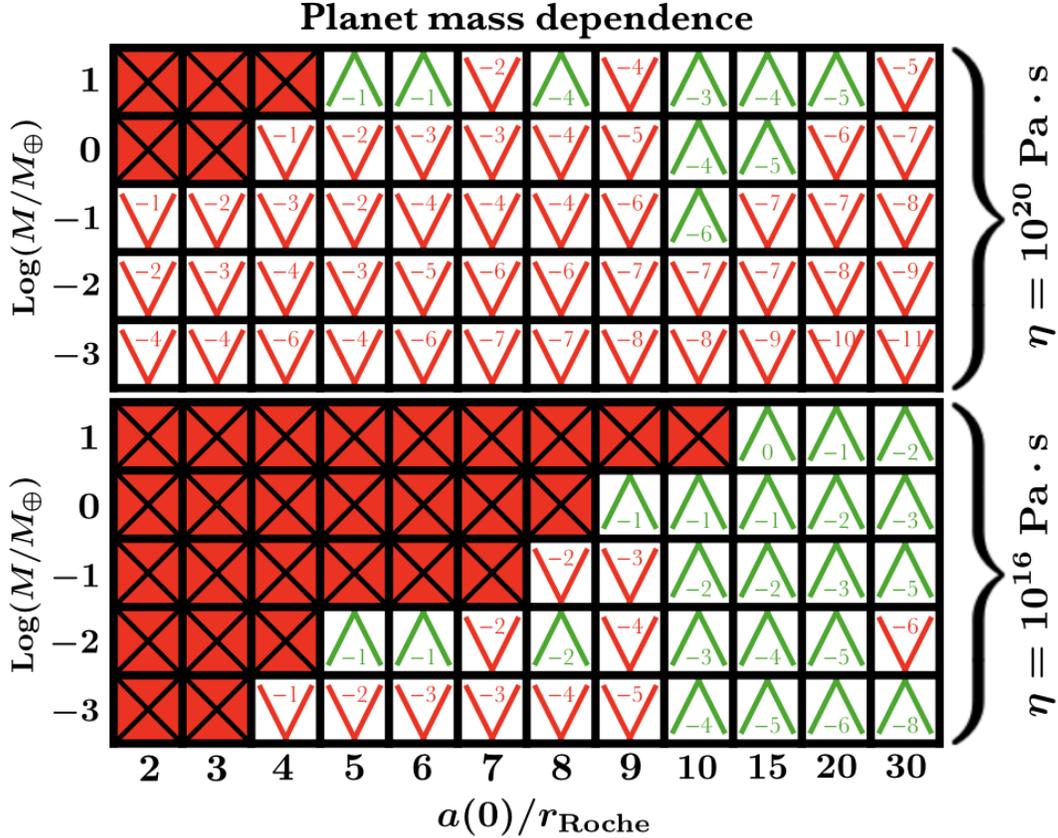}
\caption{
Same as Fig. \ref{etaplot} but for planet mass dependence, and assuming $\rho = \rho_{\oplus}$
in all cases, with $\eta = 10^{20}$ Pa$\cdot$s for the top panel and $\eta = 10^{16}$ Pa$\cdot$s for the bottom panel.
The final outcome is strongly dependent on planetary mass.
}
\label{massplot}
\end{figure*}
%%%%%%%%%%%%%%%% Figure

%%%%%%%%%%%%%%%% Figure
\begin{figure*}
\includegraphics[width=15cm]{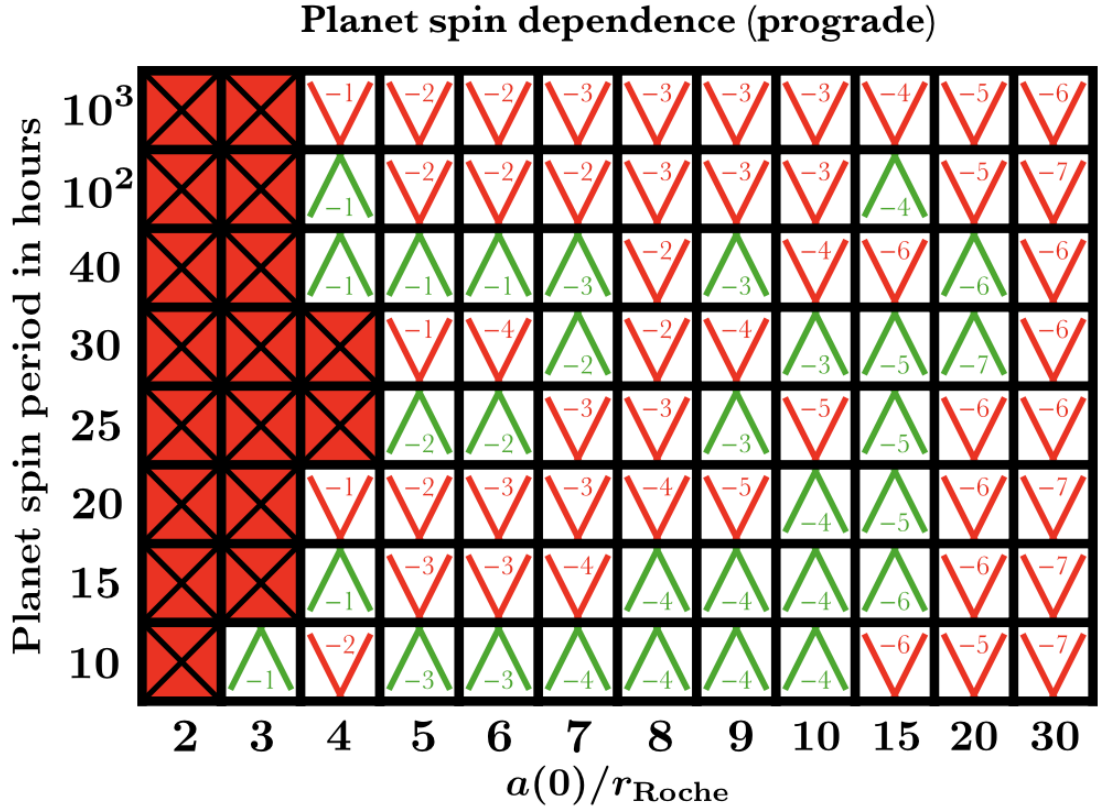}
\\
\vspace{0.5cm}
\includegraphics[width=15cm]{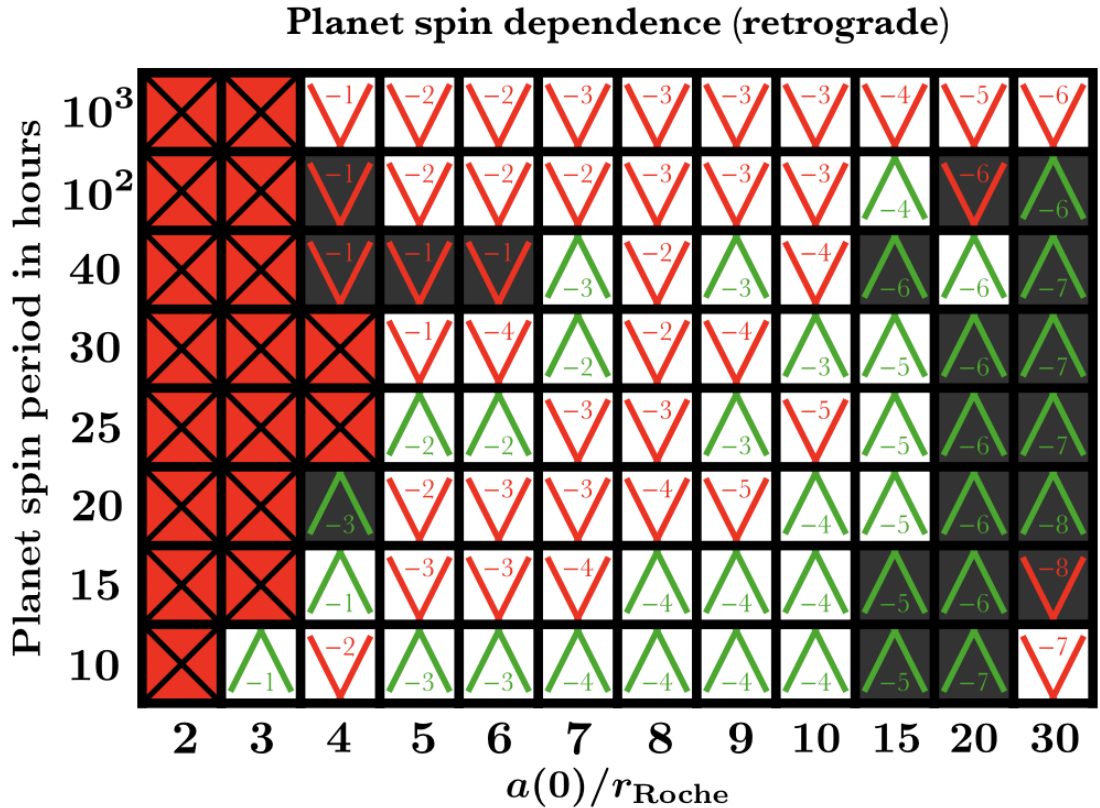}
\caption{
Same as Fig. \ref{etaplot} but for planetary spin period dependence, and assuming $\eta = 10^{20}$ Pa$\cdot$s.
The upper grid gives results for prograde planetary spins (relative to orbit orientation), and the bottom grid for 
retrograde planetary spins. The grids are 79 per cent coincident; gray squares in the bottom grid locate the discrepancies.
The final outcome has a complex dependence on planetary spin period.
}
\label{spins}
\end{figure*}
%%%%%%%%%%%%%%%% Figure

%%%%%%%%%%%%%%%% Figure
\begin{figure*}
\includegraphics[width=15cm]{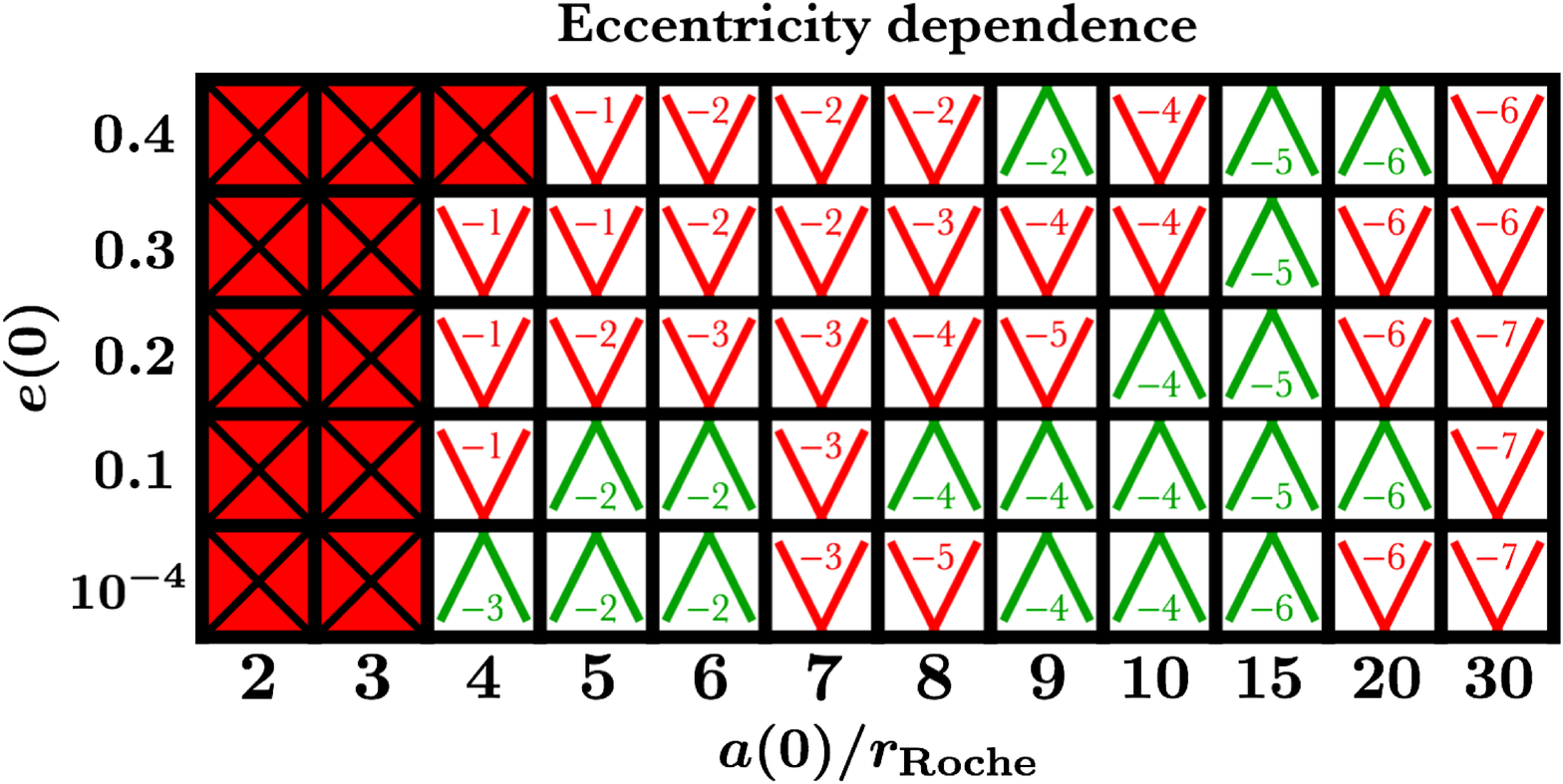}
\caption{
Same as Fig. \ref{etaplot} but for planetary eccentricity dependence, and assuming $\eta = 10^{20}$ Pa$\cdot$s.
The critical engulfment distance appears to be weakly dependent on initial eccentricity, and near-circular
orbits produce non-obvious outcomes as a function of $a(0)$.
}
\label{eccplot}
\end{figure*}
%%%%%%%%%%%%%%%% Figure

\section{Simulation results}

The simulation output of primary interest is how the planet drifts relative to the Roche radius, and whether
or not the planet reaches the Roche radius. Therefore, we report the semimajor axis drift in terms of the parameter

\begin{equation}
\sigma_a \equiv \frac{\rm final \ distance \ - \ initial \ distance}{\rm initial \ distance \ - \ Roche \ radius}.
\label{sigmaa}
\end{equation}

\noindent{}The sign of $\sigma_a$ indicates if the planet has drifted outward (positive) or inward (negative);
$\sigma_a = -100\%$ indicates entering the Roche radius (with presumed destruction subsequently).  Further, we define

\begin{equation}
\sigma_e \equiv \frac{\rm eccentricity \ change}{\rm initial \ eccentricity}
\label{sigmae}
\end{equation}

\noindent{}and

\begin{equation}
\sigma_i \equiv \frac{\rm inclination \ change}{\rm initial \ inclination}
\label{sigmai}
.
\end{equation}

\noindent{}Both $\sigma_e$ and $\sigma_i$ may be positive or negative.

Before reporting on our ensemble results, we first illustrate a couple examples of time evolution
in Figure \ref{cases}. The figure helps exhibit that the orbital evolution is not necessarily monotonic
with time, and how the final outcomes reported can be strongly 
dependent on the stopping time of the simulation.
Therefore, characterising the entire phase space within the space of one paper is challenging.
The planet can change direction due to tides, with the semimajor axis and eccentricity evolving in a non-obvious manner.
The figure also illustrates how qualitatively different behaviour can result just by decreasing the 
planet's viscosity by an order of magnitude.

That figure alone flags the danger of attempting to characterise tidal effects in an individual white
dwarf planetary system by appealing to simplified comparisons (Section 3.1) or by identifying a
particular point on a phase portrait. Nevertheless, here we do now construct phase portraits. These
are intended only for order-of-magnitude use, and to detect general trends which remain robust amidst
the complexities.

Of primary interest to us is the final value of $\sigma_a$.
We integrated the equations of motion for 100 Myr and recorded the values of $\sigma_a$
after this time has elapsed. For these simulations, we adopted the fixed initial
conditions and parameters reported in Section 4.1 ($\mathcal{J}^{\star} = 0$ (1/Pa), $i(0) = 140^{\circ}$,
$i'(0) = {10^{-4}}^{\circ}$, $\theta(0) =  \theta^{\star}(0) = 0^{\circ}$,
$(d \theta^{\star}/dt)(0) = 360^{\circ}/ (20 \ {\rm hours})$,
$M_{\star} = 0.60M_{\odot}$,
$R_{\star} = 8900$ km,
$\eta^{\star} = 10^7$ Pa$\cdot$s) and a fiducial set of initial conditions and parameters in Section 4.2
from which we vary selected parameters. This fiducial set is ($e(0) = 0.2$,
$M = M_{\oplus}$, $R = R_{\oplus}$,
$(d\theta/dt)(0) = 360^{\circ}/ (20 \ {\rm hours})$, and
$\eta = 10^{20}$ Pa$\cdot$s).

We report our results in a series of pictographs (Figs. \ref{etaplot}-\ref{eccplot}). In every
figure, we sampled 12 initial values of $a(0)$ where relevant behaviour occurs. Red crosses
indicate that $a \le r_{\rm Roche}$ at some point during the simulation, suggesting complete
destruction of the planet\footnote{We do not report when the orbital pericentre $a(1-e)$ first intersects
the Roche radius, because (i) the resulting destruction could be intermittent, and (ii) the eccentricity
is small enough within $2r_{\rm Roche}$ such that uncertainty in the Roche radius coefficient
\citep{veretal2017} would dominate the difference between orbital pericentre
and semimajor axis.}.
The other figure entries feature a red ``V'' or green caret; the former indicates the semimajor axis
has decreased ($\sigma_a < 0$) after 100 Myr, and the latter indicates that the semimajor axis has
increased ($\sigma_a > 0$) after 100 Myr. The number given inside the symbols is the order of magnitude
of the change: $-1$ refers to $10 \% \le \left| \sigma_a \right| < 100 \%$; $-3$ refers to
$0.1 \% \le \left| \sigma_a \right|< 1 \%$,
and so forth. In one case, a $0$ indicates $100 \% \le \sigma_a < 1000 \%$

The figures are ordered according to the variables which most strongly dictate the final outcome. 
The first figure, Fig. \ref{etaplot}, illustrates the importance of knowing,
estimating or being able to guess the planet's dynamic viscosity. Recall that all viscosity values in the figure are
within the ranges suggested for various components of solar system bodies.
The critical destruction distance can vary significantly, and planets which just avoid destruction
(bordering the red crosses) may be shuffled about significantly. The plot demonstrates a clear
trend against destruction as the planet viscosity increases. For the highest viscosities, even planets
initially within $2r_{\rm Roche}$ can survive.

The next figure (Fig. \ref{massplot}) reinforces the arguments presented in Section 3.5:
as planetary mass (and radius) decreases, tidal effects quickly become negligible. Similarly,
Super-Earths ($10 M_{\oplus}$) are particularly susceptible to white dwarf tides, with a (100 Myr) destruction
radius reaching out to $4 r_{\rm Roche}$ for $\eta~=~10^{20}$~Pa$\cdot$s and 
$10 r_{\rm Roche}$ for $\eta~=~10^{16}$~Pa$\cdot$s. Because asteroids are many orders of magnitude less massive than
$10^{-3} M_{\oplus}$, extrapolating from the figure indicates that spherical asteroids would be virtually
unaffected by white dwarf tidal torques, even when assuming our lower bound for dynamical viscosities.

Fig. \ref{spins} contains one grid for prograde planetary spins (top) and one for
retrograde planetary spins (bottom): differences are highlighted by gray squares. 
The grids are 79 per cent coincident with each other,
indicating that on an order-of-magnitude scale, the direction of spin makes little difference
to the final outcome. The likely reason is that the 
planet's rotation becomes quickly synchronized or pseudosynchronized. Most of the differences
occur at the highest $a(0)$ values sampled and hence are small, but there are a few important
exceptions close to the critical destruction distance.

The general
dependence of destruction and drift with the magnitude of the spin period is not obvious, at least 
from the figure.
Nevertheless, the top rows in each grid are well-ordered and perhaps suggests predictable
asymptotic behaviour as the spin period tends towards infinity.

%20/96

The final figure, Fig. \ref{eccplot}, illustrates how the critical engulfment distance is a weak function
of $e(0)$, at least for $e(0) < 0.4$. For the lowest values of $e(0)$, the results are non-obvious: the boundary
between being engulfed and remaining nearly stationary is sharp, and not adequately sampled with our
choices of $a(0)$. These results emphasise the importance of performing dedicated studies for individual systems due
to the chaotic nature of the phase space.

 \section{Summary}

Exo-asteroids are already observed orbiting two white dwarfs in real time, and almost every known exo-planet currently orbits a star that will become a white dwarf. Planets which then survive to the white dwarf phase play a crucial role in frequently shepherding asteroids and their observable detritus onto white dwarf atmospheres, even if the planets themselves lie just outside of the narrow range of detectability. Further, planets themselves may occasionally shower a white dwarf with metal constituents through post-impact crater ejecta and when the planetary orbit grazes the star's Roche radius \citep{vergae2015,broetal2017}.

Here, we undertook one of the first dedicated studies of tides in a two-body system comprising a solid planet and a white dwarf. We adopted the Fourier-mode tidal formalism of \cite{bouefr2019}, which provides complete equations (\ref{aevol}--\ref{ipevol}) for the secular evolution of the orbital semimajor axis, eccentricity and inclination with frequency dependencies on the quality functions. By combining these equations with secular parts of the tidal torque from \cite{efroimsky2012a} and \cite{makarov2012}, and typical physical parameters for the two bodies (Section 4.1), we generated a computational framework for future, more detailed consideration of individual systems.

A broad sweep of phase space has revealed the following trends:

{\bf (i)} Massive Super-Earths are more easily destroyed than low-mass planets. 

{\bf (ii)} Planetary survival is boosted with higher viscosities.  

{\bf (iii)} Orbital evolution is in general non-monotonic, and cannot usually be described by a straightforward comparison of spin period versus orbital period. 

{\bf (iv)} The boundary between destruction and survival appears to be chaotic.

{\bf (v)} Because the magnitude of the stellar tides scales as the mass of the perturber, the orbital dynamics of the asteroids in the WD 1145+017 and SDSS J1228+1040 systems are unaffected by stellar tides.

{\bf (vi)} The relatively small range of white dwarf physical parameters, as compared to those of main-sequence stars, helps constrain variable explorations and ensure that semimajor axis, planetary mass, planetary viscosity and planetary spin rates are the four key variables which influence the evolution. 

{\bf (vii)} Conservatively, even for low values of the planetary viscosity, planets which do not achieve a semimajor axis of under
about $10r_{\rm Roche}$ will survive for at least 100 Myr of white-dwarf cooling. 

{\bf (viii)} The planet's evolution is largely independent of the direction but not magnitude of the spin of the planet.

{\bf (ix)} Despite the varied ways in which the orbit can be stretched (Figure \ref{cases}), the critical engulfment distance is largely independent of the orbital eccentricity when it takes on low and moderate values.

\section*{Acknowledgements}

All authors appreciate the helpful comments from the MNRAS referee and 
the four internal United States Naval Observatory 
(USNO) referees Paul Barrett, Christopher Dieck, George Kaplan and Robert Zavala.
DV further thanks the USNO for its repeated hospitality for over a decade, and the Landessternwarte, Zentrum f\"{u}r Astronomie der Universit\"{a}t Heidelberg for a productive visit, ultimately kickstarting this work.  DV also gratefully acknowledges the support of the STFC via an Ernest Rutherford Fellowship (grant ST/P003850/1).
ME would like to thank Alexander Getling for a valuable consultation on the stellar viscosity.
VW and SR further acknowledge support by the DFG Priority Program SPP 1992 Exploring the Diversity 
of Extrasolar Planets (RE 2694/5-1). PET has received funding from the European Research Council under 
the European Union's Horizon 2020 research and innovation programme n. 677706 (WD3D).

 %  \pagebreak 

\appendix

\onecolumn

 \section{Parameter Tables}

 \begin{table*}
 \centering
 \begin{minipage}{180mm}
  \centering
  \caption{Unstylised and stylised Roman variables used in this paper, with notation that assumes a stellar primary and planetary secondary. No functional dependencies are shown.  Variables with a $\star$ subscript or superscript refer to a stellar physical property.}
  \label{description}
  \begin{tabular}{@{}llll@{}}
  \hline
   Variable & Explanation & Units & Reference \\
 \hline
 $a$ & Semimajor axis of the mutual orbit & length & equation (\ref{aevol})  \\[2pt]
% $\mathfrak{a}$ & Longest axis of planet &  \\[2pt]
 %$A$ & Minimial moment of inertia of the planet &  \\[2pt]
 %$A^{\star}$ & Minimial moment of inertia of the star &  \\[2pt]
% $\mathfrak{b}$ & Middle axis of planet &  \\[2pt]
 %$B$ & Middle moment of inertia of the planet &  \\[2pt]
 %$B^{\star}$ & Middle moment of inertia of the star &  \\[2pt]
 $\mathcal{B}$ & Auxiliary variable  & 1/pressure  & equation (\ref{planB}) \\[2pt]
 $\mathcal{B}^{\star}$ & Auxiliary variable  &  1/pressure &  equation (\ref{starB})  \\[2pt]
% $\mathfrak{c}$ & Shortest axis of planet &  \\[2pt]
 $d_{\star}$ & Typical lengthscale of stellar convective flows & distance & equation (\ref{granule})  \\[2pt]
 $e$ & Eccentricity of the mutual orbit & dimensionless  & equation (\ref{eevol})   \\[2pt]
 $F$ & Inclination function & dimensionless  & equation (\ref{incfunc})  \\[2pt]
 $g$ & Acceleration due to gravity on planet surface & length/time$^2$  & \\[2pt]
 $g_{\star}$ & Acceleration due to gravity on stellar surface &  length/time$^2$ & \\[2pt]
 $G$ & Eccentricity function & dimensionless  & equation (\ref{eccfunc})  \\[2pt]
 $\mathcal{G}$ & Gravitational constant & length$^3$/(mass$\times$time$^2$)  & \\[2pt]
 $i$ & Inclination of the mutual orbit with respect to the planet's equator & angle & equation (\ref{ievol}) \\[2pt]
 $i^{\prime}$ & Inclination of the mutual orbit with respect to the star's equator & angle & equation (\ref{ipevol}) \\[2pt]
% $j$ & Integer used as summation index for the tidal torque & \\[2pt]
 $J$ & Bessel function of the first kind & dimensionless & \\[2pt]
% $\mathfrak{J}_{2}$ & Second gravitational moment of planet & \\[2pt]
% $\mathfrak{J}_{2}^{\star}$ & Second gravitational moment of star & \\[2pt]
 $\mathcal{J}$ & Compliance of the planet & 1/pressure  & \\[2pt]
 $\mathcal{J}^{\star}$ & Compliance of the star & 1/pressure  &  \\[2pt]
 $k$ & Love number of the planet & angle/time &  \\[2pt]
 $k^{\star}$ & Love number of the star & angle/time & \\[2pt]
 $l$ & Positive integer used as Love number index (``degree'') & dimensionless &   \\[2pt]
 $L_{\star}$ & Luminosity of star & power & Table 1  \\[2pt]
 $m$ & Integer used as tidal mode index and inclination function index (``order'') & dimensionless &   \\[2pt]
 $M$ & Mass of planet & mass &   \\[2pt]
 $M_{\star}$ & Mass of star &  mass & \\[2pt]
% $\mathcal{M}$ & Mean anomaly of planet along mutual orbit &   \\[2pt]
 $n$ & Anomalistic mean motion  &  angle/time &   \\[2pt]
% $o_{e}$ & Order of eccentricity & \\[2pt]
% $o_{i}$ & Order of sum of powers of sines and cosines of inclination & \\[2pt]
 $p$ & Integer for tidal mode index, eccentricity function index, \& inclination function index & dimensionless &  \\[2pt]
 $q$ & Integer for tidal mode index and eccentricity function index & dimensionless &  \\[2pt]
 $R$ & Radius of planet & distance &  \\[2pt]
 $R_{\star}$ & Radius of star & distance & \\[2pt]
 $s$ & Integer used as summation index for the eccentricity function & dimensionless &   \\[2pt]
 $s_1$ & Integer used as summation index limit for the eccentricity function & dimensionless & \\[2pt]
 $t$ & time &  time & \\[2pt]
 $T_{\rm eff}^{\star}$ & Effective temperature of star & temperature & Table 1  \\[2pt]
 $\mathcal{T}$ & Torque & mass$\times$length$^2$/time$^2$  & \\[2pt]
 $u$ & Integer used as summation index for the eccentricity and inclination functions & dimensionless &  \\[2pt]
% $u_1$ & Whole number used as summation index limit for the inclination function &   \\[2pt]
 $v_{\star}$ & Typical speed of stellar supergranules & length/time & equation (\ref{granule})   \\[2pt]
\hline
\end{tabular}
\end{minipage}
\end{table*}

\begin{table*}
 \centering
 \begin{minipage}{180mm}
  \centering
  \caption{As in Table 1, but with Greek variables.}
  \label{description}
  \begin{tabular}{@{}llll@{}}
  \hline
   Variable & Explanation & Units & Reference \\
  \hline
 $\delta$ & Delta function & dimensionless & \\[2pt]
 $\epsilon$ & Phase lag of the planet & dimensionless  & \\[2pt]
 $\epsilon^{\star}$ & Phase lag of the star & dimensionless & \\[2pt]
 $\eta$ & Dynamic viscosity of the planet & pressure$\times$time &  \\[2pt]
 $\eta^{\star}$ & Dynamic viscosity of the star & pressure$\times$time &  equation (\ref{granule})  \\[2pt]
 $\theta$ & Rotation angle about the instantaneous shortest axis of planet & angle & equation (\ref{thetaevol})  \\[2pt]
 $\theta^{\star}$ & Rotation angle about the instantaneous shortest axis of star  & angle &   \\[2pt]
 $\xi$ & Coefficient for planet's moment of inertia ($=\frac{2}{5}$ for a homogeneous sphere) & dimensionless & \\[2pt]
 $\xi^{\star}$ & Coefficient for star's moment of inertia ($=\frac{2}{5}$ for a homogeneous sphere) & dimensionless & \\[2pt]
 $\rho$ & Density of planet & mass/length$^3$ & \\[2pt]
 $\rho^{\star}$ & Density of star &  mass/length$^3$ & \\[2pt]
 $\sigma_a$ & Semimajor axis decrease divided by the initial distance to the Roche radius & dimensionless & equation (\ref{sigmaa})  \\[2pt]
 $\sigma_e$ & Eccentricity decrease divided by the initial eccentricity & dimensionless & equation (\ref{sigmae})  \\[2pt]
 $\sigma_i$ & Inclination decrease divided by the initial inclination & dimensionless  & equation (\ref{sigmai})  \\[2pt]
%$\sigma'$ & Argument of pericentre of orbit with respect to the planet &   \\[2pt]
 $\chi_{lmpq}$ & Positive definite physical forcing frequency in the planet & angle/time & equation (\ref{chiplan})  \\[2pt]
 $\chi^{\star}_{lmpq}$ & Positive definite physical forcing frequency in the star & angle/time  & equation (\ref{chistar})  \\[2pt]
 $\omega_{lmpq}$ & Fourier tidal mode in the planet & angle/time & equation (\ref{omegaplan})  \\[2pt]
 $\omega^{\star}_{lmpq}$ & Fourier tidal mode in the star & angle/time & equation (\ref{omegastar})  \\[2pt]
 %$\Omega$ & Longitude of ascending node of orbit with respect to the star &   \\[2pt]
 %$\Omega'$ & Longitude of ascending node of orbit with respect to the planet &   \\[2pt]
\hline
 $k_l \sin{\epsilon_l}$ & Quality function of the planet & angle/time & equation (\ref{qualplan}) \\[2pt]
 $k^{\star}_l\sin{\epsilon^{\star}_l}$ & Quality function of the star & angle/time & equation (\ref{qualstar})  \\[2pt]
\hline
\end{tabular}
\end{minipage}
\end{table*}

\pagebreak

\section{Eccentricity and inclination functions}

The eccentricity and inclination functions are utilised in the equations of motion, and may
be precomputed regardless of the physical setup. Doing so helps speed up the integrations.

\subsection{Eccentricity function}

We utilise the useful form of the eccentricity function $G_{lpq}(e)$ which is given
in the Appendix of \cite{celetal2017}. Their
formulae are based on those from \cite{giacaglia1976}. Transforming their notation
into our formalism yields the following results for the eccentricity function

\begin{equation}
  G_{lpq}(e) =
  \left[1 + \left( \frac{e}{1 + \sqrt{1-e^2} }\right)^2 \right]^{l}
  \sum_{s=0}^{s_1} \sum_{u=0}^{u_1}
  \left(
       {\begin{array}{cc}
   2p-2l \\
   s \\
       \end{array} }
  \right)
  \left(
       {\begin{array}{cc}
   -2p \\
   u \\
       \end{array} }
  \right)
  \left( -\frac{e}{1 + \sqrt{1-e^2} }\right)^{s+u}
  J_{q-s+u}\left(\left(l-2p+q\right)e\right)
   \label{eccfunc}
\end{equation}

\noindent{}where

\[
s_1 = 2p-2l, \ \ \ {\rm if} \ p-l \ge 0
\]
\begin{equation}
s_1 = \infty, \ \ \ \ \ \ \ \ \ {\rm if} \ p-l < 0
\end{equation}

\noindent{}and

\[
u_1 = -2p, \ \ \ \ {\rm if} \ p \le 0
\]
\begin{equation}
u_1 = \infty, \ \ \ \ \ \ {\rm if} \ p > 0
.
\end{equation}

\noindent{}Note that because $0 \le p \le l$, always here $s_1 = 2p-2l$ and
$u_1 = \infty$. This last relation creates an additional parameter in numerical computations
which must be explored in order to achieve the desired accuracy.

%Fig. \ref{G22} suggests that $G$ converges quickly for $e=0.8$ and for a sufficiently high values of $u_1$ and $q$. We sampled two different values $u_1 = 30$ (blue curve) and $u_1 = 50$ (yellow curve) on each plot. Although the curves on the same plot appear nearly indistinguisible, given the range $-200 \le q \le 200$,  the greatest per cent error in the solutions is $-2.8 \%$ for $G_{22q}$ (at $q=200$) and  $-11.8 \%$  for $G_{44q}$ (at $q=-200$).

%%%%%%%%%%%%%%%% Figure
%\begin{figure*}
%\includegraphics[width=8cm]{G22q.eps}
%\includegraphics[width=8cm]{G44q.eps}
%\caption{
%$G_{22q}$ (left panel) and $G_{44q}$ (right panel) as a function of $q$ assuming $u_1 = 30$ (blue curve) and $u_1 = 50$ (yellow curve) for $e=0.8$.
%}
%\label{G22}
%\end{figure*}
%%%%%%%%%%%%%%%% Figure

%%%%%%%%%%%%%%%% Figure
%\begin{figure}
%\includegraphics[width=8cm]{G33q.eps}
%\caption{
%$G_{33q}$ as a function of $q$ assuming $u_1 = 30$ (blue curve) and $u_1 = 50$ (yellow curve) for $e=0.8$.
%}
%\label{G33}
%\end{figure}
%%%%%%%%%%%%%%%% Figure

%%%%%%%%%%%%%%%% Figure
%\begin{figure}
%\includegraphics[width=8cm]{G44q.eps}
%\caption{
%$G_{44q}$ as a function of $q$ assuming $u_1 = 30$ (blue curve) and $u_1 = 50$ (yellow curve) for $e=0.8$.
%}
%\label{G44}
%\end{figure}
%%%%%%%%%%%%%%%% Figure

\subsection{Inclination function}

We use the inclination function given in equation 4 of \cite{goowag2008}, which corrects some
sign errors from the exposition in \cite{kaula1962} and conveniently does not include integer
parts. Translating the \cite{goowag2008} variables into ours yields

\begin{equation}
  F_{lmp}(i) =
  \frac{\left(l+m\right)!}{2^l p! \left(l-p\right)!}
  \sum_{u={\rm max}\left(0, \ l-2p-m\right)}^{{\rm min}\left(2l-2p, \ l-m\right)}
  \left(-1\right)^{u}
   \left(
      {\begin{array}{cc}
   2l-2p \\
   u \\
       \end{array} }
   \right)
    \left(
      {\begin{array}{cc}
   2p \\
   l-m-u \\
       \end{array} }
    \right)
    \cos^{3l-m-2p-2u}\left(\frac{i}{2}\right)
    \sin^{m-l+2p+2u}\left(\frac{i}{2}\right)
.
\label{incfunc}
\end{equation}

 \section{Equations of motion\label{app}}

The following equations of motion (equations \ref{aevol}-\ref{tid1star})
include the integers $\,l,\,m,\,p$, and $q$ explicitly through summation limits; note that $n$ here
refers to mean motion (specifically, the anomalistic mean motion) and is not an index.

 Known in tidal investigations as the ``degree'', the positive integer $\,l\,$ in particular sets the scene: (a) it represents the index for the first summation which appears in these equations of motion, and (b) it establishes the maximum possible values of $m$ and $p$.

 For point (a), the truncation of the tidal expansions by setting $l=2$ (the quadrupole approximation) represents a very common approximation for the evolution of the orbital elements and the spin. The suitability of the quadrupole approximation is discussed in \cite{bouefr2019}. Although for highly eccentric orbits around white dwarfs this approximation might not be adequate, it remains sufficient for the eccentricities considered here ($<$0.40).

For point (b), because $l=2$, then $p=0,1,2$ and $m=0,1,2$. The only index which then remains a free parameter is $q$.  Consequently, the range of $q$ must be chosen to encompass all of the eccentricity terms of a given order. Although a typical approximation is to assume that $\left|q\right|$ is equal to the highest power of the eccentricity sampled, actually the distribution of $q$ which must be considered to achieve a given accuracy is strongly asymmetric (\citealt*{noyetal2014} use $-1 \le q \le 7$). We did not employ the typical approximation for $q$, nor consider the concept of order, which can introduce pitfalls in series expansions \citep{veras2007}. We created fits to $G_{lpq}(e)$ for $e<0.4$ by choosing, quite conservatively, $-28 \le q \le 43$. However, preliminary numerical testing indicated that we need only choose $-7 \le q \le 7$ for our primary integrations; this range of values allowed us computationally to sample several regions of phase space.

All of the equations in this section include tidal contributions from both the star
 and planet, and are not truncated in any way.

 \subsection{Equations of orbital motion}

The secular semimajor axis evolution of the mutual orbit is given by \cite{bouefr2019} as

\[
\left\langle
\left\langle
\frac{da}{dt}
\right\rangle
\right\rangle
=
-2 a n \sum_{l=2}^{\infty} \sum_{m=0}^{l}
       \frac{\left(l-m\right)!}{\left(l+m\right)!} \left(2-\delta_{0m}\right)
       \sum_{p=0}^{l} \sum_{q=-\infty}^{\infty} G_{lpq}^2(e) \left(l-2p+q\right)
\]

\begin{equation}
  \ \ \ \ \
  \times
  \left[
  \left(\frac{R}{a}\right)^{2l+1} \frac{M_{\star}}{M} F_{lmp}^2(i)
           k_{l}(\omega_{lmpq}) \sin{\left[\epsilon_{l}(\omega_{lmpq})\right]}
 +\left(\frac{R_{\star}}{a}\right)^{2l+1} \frac{M}{M_{\star}} F_{lmp}^2(i')
           k_{l}^{\star}(\omega_{lmpq}^{\star}) \sin{\left[\epsilon_{l}^{\star}(\omega_{lmpq}^{\star})\right]}
  \right]
.
\label{aevol}
\end{equation}
\noindent{}where the double averaging indicates averaging over both one orbital cycle and one cycle of the apsidal precession.

The secular eccentricity evolution of the mutual orbit is given by \cite{bouefr2019} as

\[
\left\langle
\left\langle
\frac{de}{dt}
\right\rangle
\right\rangle
=
-\frac{1-e^2}{e}   n  \sum_{l=2}^{\infty} \sum_{m=0}^{l}
       \frac{\left(l-m\right)!}{\left(l+m\right)!} \left(2-\delta_{0m}\right)
       \sum_{p=0}^{l} \sum_{q=-\infty}^{\infty} G_{lpq}^2(e) \left(l-2p+q\right)
\]

\[
  \ \ \ \ \
\times \left[
  \left(\frac{R}{a}\right)^{2l+1} \frac{M_{\star}}{M} F_{lmp}^2(i)
           k_{l}(\omega_{lmpq}) \sin{\left[\epsilon_{l}(\omega_{lmpq})\right]}
 +\left(\frac{R_{\star}}{a}\right)^{2l+1} \frac{M}{M_{\star}} F_{lmp}^2(i')
           k_{l}^{\star}(\omega_{lmpq}^{\star}) \sin{\left[\epsilon_{l}^{\star}(\omega_{lmpq}^{\star})\right]}
\right]
\]

\[
  \ \ \ \ \
+
\frac{\sqrt{1-e^2}}{e}n \sum_{l=2}^{\infty} \sum_{m=0}^{l}
       \frac{\left(l-m\right)!}{\left(l+m\right)!} \left(2-\delta_{0m}\right)
       \sum_{p=0}^{l} \sum_{q=-\infty}^{\infty} G_{lpq}^2(e) \left(l-2p\right)
\]

\begin{equation}
    \ \ \ \ \
\times \left[
  \left(\frac{R}{a}\right)^{2l+1} \frac{M_{\star}}{M} F_{lmp}^2(i)
           k_{l}(\omega_{lmpq}) \sin{\left[\epsilon_{l}(\omega_{lmpq})\right]}
 +\left(\frac{R_{\star}}{a}\right)^{2l+1} \frac{M}{M_{\star}} F_{lmp}^2(i')
           k_{l}^{\star}(\omega_{lmpq}^{\star}) \sin{\left[\epsilon_{l}^{\star}(\omega_{lmpq}^{\star})\right]}
\right]
.
\label{eevol}
\end{equation}

\noindent{}The secular inclination evolution of the mutual orbit with respect to the planetary equator is given by \cite{bouefr2019} as

\[
\left\langle \left\langle
\frac{di}{dt}
\right\rangle \right\rangle
=
  -\;\frac{n}{\sqrt{1-e^2}}\;\frac{{M_{\star}}}{M\;}
 \sum_{l=2}^{\infty}\left(\frac{R}{a}\right)^{\textstyle{^{2l+1}}}\sum_{m=0}^{l}\frac{(l - m)!}{(l + m)!}\left(2- \delta_{0m}\right)
 %  \quad\qquad\qquad\qquad\\ \label{}\\ \nonumber
\]

\[
 \ \ \ \ \ \ \ \ \ \times \sum_{p=0}^{l}
 \frac{(l-2p)\cos i-m}{\sin i}\,F^{\,2}_{lmp}(i)\sum_{q=-\infty}^{\infty}G^{\,2}_{lpq}(e)\,k_{l}(\omega_{lmpq}) \sin{\left[\epsilon_{l}(\omega_{lmpq})\right]}
\]

\[
\ \ \ \ \ \ \ \ \
+
\left\langle
\frac{d\theta}{dt}
\right\rangle^{-1}
 \;\frac{n^2 a^2}{\xi M R^2}\;\frac{M_{\star}^2}{M + M_{\star}}
 \sum_{l=2}^\infty \left(\frac{R}{a}\right)^{\textstyle{^{2l+1}}}\sum_{m=0}^{l}\frac{(l-m)!}{(l+m)!}\left(2-\delta_{0m}\right)
\]

\begin{equation}
\ \ \ \ \ \ \ \ \ \times \sum_{p=0}^{l} \frac{m\cos i-(l-2p)}{\sin i}\,F^{\,2}_{lmp}(i)
 \sum_{q=-\infty}^{\infty}G^{\,2}_{lpq}(e)\,k_{l}(\omega_{lmpq}) \sin{\left[\epsilon_{l}(\omega_{lmpq})\right]}
\label{ievol}
.
\end{equation}

\noindent{}With respect to the stellar equator, we instead have

\[
\left\langle \left\langle
\frac{di'}{dt}
\right\rangle \right\rangle
=
  -\;\frac{n}{\sqrt{1-e^2}}\;\frac{{M}}{M_{\star}}
 \sum_{l=2}^{\infty}\left(\frac{R_{\star}}{a}\right)^{\textstyle{^{2l+1}}}\sum_{m=0}^{l}\frac{(l - m)!}{(l + m)!}\left(2- \delta_{0m}\right)
 %  \quad\qquad\qquad\qquad\\ \label{}\\ \nonumber
\]

\[
 \ \ \ \ \ \ \ \ \ \times \sum_{p=0}^{l}
 \frac{(l-2p)\cos i'-m}{\sin i'}\,F^{\,2}_{lmp}(i')\sum_{q=-\infty}^{\infty}G^{\,2}_{lpq}(e)\,k_{l}^{\star}(\omega_{lmpq}^{\star}) \sin{\left[\epsilon_{l}^{\star}(\omega_{lmpq}^{\star})\right]}
\]

\[
\ \ \ \ \ \ \ \ \
+
\left\langle
\frac{d\theta^{\star}}{dt}
\right\rangle^{-1}
 \;\frac{n^2 a^2}{\xi^{\star} M_{\star} R_{\star}^2}\;\frac{M^2}{M + M_{\star}}
 \sum_{l=2}^\infty \left(\frac{R_{\star}}{a}\right)^{\textstyle{^{2l+1}}}\sum_{m=0}^{l}\frac{(l-m)!}{(l+m)!}\left(2-\delta_{0m}\right)
\]

\begin{equation}
\ \ \ \ \ \ \ \ \ \times \sum_{p=0}^{l} \frac{m\cos i'-(l-2p)}{\sin i'}\,F^{\,2}_{lmp}(i')
 \sum_{q=-\infty}^{\infty}G^{\,2}_{lpq}(e)\,k_{l}^{\star}(\omega_{lmpq}^{\star}) \sin{\left[\epsilon_{l}^{\star}(\omega_{lmpq}^{\star})\right]}
\label{ipevol}
.
\end{equation}

\subsection{Equations of spin motion}

The secular spin evolutions of the planet and star are given by Eq. 1 of \cite{makarov2012} as

\begin{equation}
\left\langle\left\langle
\frac{d^2\theta}{dt^2}
\right\rangle\right\rangle
= \frac{ \left\langle  \mathcal{T}_{z}^{({\rm tri})} \right\rangle
  +
              \left\langle\left\langle \mathcal{T}_{z}^{({\rm tide})} \right\rangle\right\rangle  }{\xi M R^2},
\label{thetaevol}
\end{equation}

\begin{equation}
\left\langle\left\langle
\frac{d^2\theta^{\star}}{dt^2}
\right\rangle\right\rangle
= \frac{\left\langle  \mathcal{T}_{z,\star}^{({\rm tri})} \right\rangle
+
 \left\langle\left\langle  \mathcal{T}_{z,\star}^{({\rm tide})} \right\rangle\right\rangle }{\xi^{\star} M_{\star} R_{\star}^2},
\label{thetaevolstar}
\end{equation}

\noindent{}and the secular part of the polar component of the tidal torque is given by equation (109) of \cite{efroimsky2012a} as
%\noindent{}and the $l=2$ component of the tidal torque is, from equation 11a of the same paper,

\begin{equation}
\left\langle\left\langle
  \mathcal{T}_{z}^{({\rm tide})}
\right\rangle\right\rangle
=
2 G M^2 \sum_{l=2}^{\infty} \frac{R^{2l+1}}{a^{2l+2}}
                \sum_{m=1}^l \frac{\left(l-m\right)!}{\left(l+m\right)!} m
                \sum_{p=0}^l F_{lmp}^2(i)
                \sum_{q=-\infty}^{\infty} G_{lpq}^2(e) k_l \sin{\epsilon_{l}\left(\omega_{lmpq} \right)}
,
\label{tid1}
\end{equation}

\begin{equation}
\left\langle\left\langle
  \mathcal{T}_{z,\star}^{({\rm tide})}
\right\rangle\right\rangle
=
2 G M_{\star}^2 \sum_{l=2}^{\infty} \frac{R_{\star}^{2l+1}}{a^{2l+2}}
                \sum_{m=1}^l \frac{\left(l-m\right)!}{\left(l+m\right)!} m
                \sum_{p=0}^l F_{lmp}^2(i')
                \sum_{q=-\infty}^{\infty} G_{lpq}^2(e) k_{l}^{\star} \sin{\epsilon_{l}^{\star}\left(\omega_{lmpq}^{\star} \right)}
.
\label{tid1star}
\end{equation}

 \noindent{}A detailed derivation of the polar component of the permanent-triaxiality-generated torque is provided in \citet[Appendix D]{frouard2017}. Equation (116e) of that paper reveals that \,{\it{under steady rotation}}\, and, importantly, {\it{outside of a spin-orbit resonance}}\,, the secular part of this torque vanishes. In realistic situations, the equation further reveals that when the apsidal precession is much slower than the orbital motion, then orbital averaging is sufficient to nullify the torque. For these reasons, we equip these secular parts with only one pair of angular brackets:
\begin{equation}
\left\langle
 \mathcal{T}_{z}^{({\rm tri})}
\right\rangle
=
0,
\label{tri}
\end{equation}

\begin{equation}
\left\langle
  \mathcal{T}_{z,\star}^{({\rm tri})}
\right\rangle
=
0.
\label{tristar}
\end{equation}
Both these equalities no longer imply that a rotator is captured in a spin-orbit resonance. In such a resonance, the triaxiality-generated torque is leading for non-liquid bodies (much larger than the tidal torque) and becomes the driver of longitudinal libration.

%\noindent{}When averaged, this tidal torque becomes (equation 11b of \citealt*{noyetal2014})

%\begin{equation}
 % \left\langle \mathcal{T}_{z}^{({\rm tide})} \right\rangle_{l=2} =
 % \frac{3}{2} \frac{\mathcal{G}M_{\star}^2}{a} \left(\frac{R}{a}\right)^5
 % \sum_{q=-1}^{7} G_{20q}^2(e)
 % k_{2}^{\star} \left(\omega_{220q}^{\star}\right) \sin{\left|\epsilon_{2}^{\star}(\omega_{220q}^{\star})\right|}
 % {\rm Sgn}\left(\omega_{220q}^{\star}\right)
%  + \mathcal{O}\left(e^8\epsilon^{\star}\right) + \mathcal{O}\left(i^2\epsilon^{\star}\right)
%\label{tid2}
%,
%\end{equation}

%\begin{equation}
% \left\langle \mathcal{T}_{z,\star}^{({\rm tide})} \right\rangle_{l=2} =
%  \frac{3}{2} \frac{\mathcal{G}M^2}{a} \left(\frac{R_{\star}}{a}\right)^5
%  \sum_{q=-1}^{7} G_{20q}^2(e)
%  k_{2} \left(\omega_{220q}\right) \sin{\left|\epsilon_{2}(\omega_{220q})\right|}
%  {\rm Sgn}\left(\omega_{220q}\right)
%  + \mathcal{O}\left(e^8\epsilon\right) + \mathcal{O}\left(i'^2\epsilon\right)
%\label{tid2star}
%.
%\end{equation}

\label{lastpage}
\end{document}